\documentclass[aps,pre,twocolumn,showpacs,amsfonts,amssymb,floatfix,superscriptaddress]{revtex4-1}

\usepackage{amsfonts}
\usepackage{amsmath}
\usepackage{amssymb}
\usepackage{bm}
\usepackage{psfrag}
\usepackage{graphicx}

\begin{document}


\title{Randomly Evolving Idiotypic Networks:\\Structural Properties
  and Architecture}



\author{Holger Schmidtchen}
\affiliation{Institut f\"ur Theoretische Physik, Universit\"at
  Leipzig, POB~100~920, D-04009~Leipzig, Germany}
\affiliation{International Max Planck Research School Mathematics in the
  Sciences,\\Inselstra\ss e 22, D-04103 Leipzig, Germany}

\author{Mario Th\"une}
\affiliation{Institut f\"ur Theoretische Physik, Universit\"at
  Leipzig, POB~100~920, D-04009~Leipzig, Germany}

\author{Ulrich Behn}
\email[]{ulrich.behn@itp.uni-leipzig.de}
\affiliation{Institut f\"ur Theoretische Physik, Universit\"at
  Leipzig, POB~100~920, D-04009~Leipzig, Germany}
\affiliation{International Max Planck Research School Mathematics in the
  Sciences,\\Inselstra\ss e 22, D-04103 Leipzig, Germany}


\date{\today}

\begin{abstract}

We consider a minimalistic dynamic model of the idiotypic network of B-lymphocytes. A network node represents a population of B-lymphocytes of the same specificity (idiotype), which is encoded by a bitstring. The links of the network connect nodes with complementary and nearly complementary bitstrings, allowing for a few mismatches. A node is occupied if a lymphocyte clone of the corresponding idiotype exists, otherwise it is empty. There is a continuous influx of new B-lymphocytes of random idiotype from the bone marrow. B-lymphocytes are stimulated by cross-linking their receptors with complementary structures.
If there are too many complementary structures, steric hindrance prevents cross-linking. Stimulated cells proliferate and secrete antibodies of the same idiotype as their receptors, unstimulated lymphocytes die.

Depending on few parameters, the autonomous system evolves randomly towards patterns of highly organized architecture, where the nodes can be classified into groups according to their statistical properties. We observe and describe analytically the building principles of these patterns, which allow to calculate number and size of the node groups and the number of links between them. The architecture of all patterns observed so far in simulations can be explained this way. A tool for real-time pattern identification is proposed.

\end{abstract}


\maketitle

\section{Introduction\label{sec:introduction}}


B-lymphocytes play a crucial role in the adaptive immune system. 
They express Y-shaped receptor molecules, antibodies, on their surface. Antibodies have very specific binding sites ({\em idiotopes}) which determine their {\em idiotype}. All receptors of a given B-cell have the same idiotype. B-cells are stimulated to proliferate if their receptors are cross-linked by structures which are complementary to the idiotopes, e.g. by foreign antigen. Stimulated B-cells thus survive whereas unstimulated B-cells die, this process is called clonal selection \cite{Burnet59}.

After a few generations, stimulated B-cells differentiate to plasma cells which secrete soluble antibody molecules of the same idiotype, which may bind to complementary sites on antigen and mark them for further processing.

B-lymphocytes of different idiotype are continuously produced in the bone marrow in a remarkable diversity. The variety of the potential idiotype repertoire, created by somatic reshuffling of gene segments and mutations \cite{Tonegawa83}, was combinatorially estimated \cite{BM88} to exceed most likely $10^{10}$.

Complementary structures may be found on antigen but could also be situated on other antibodies of complementary idiotype. B-lymphocytes can stimulate each other and, thus, they form a functional network, the {\em idiotypic network} \cite{Jerne74,*Jerne84}.

Jerne's concept of the idiotypic network explains in a natural way the diversity of the expressed  idiotype repertoire and the autonomous dynamics of an immune system not exposed to foreign antigen. It provides a mechanism of immunological memory. Imagine that an antigen $Ag$ is recognized by an antibody $Ab_1$. Thus, the clone of $Ab_1$ expands and possibly meets another clone of complementary idiotype $Ab_2$. Both mutually stimulate each other and they persist even after $Ag$ has disappeared. $Ab_2$ is structurally similar to $Ag$ and can be considered as internal image of $Ag$. Furthermore, the network is thought to control autoreactive clones. All these issues are beyond the concept of clonal selection. 

The network paradigm got an immediate enthusiastic response and idiotypic interactions were considered as the major regulating mechanism of the immune system. However, the rapid progress of molecular immunology, difficulties in the direct experimental verification, and the discovery of other regulating mechanisms let the interest of experimental immunologists decay. Yet, for system biologists the network paradigm always remained attractive. Several aspects of the original concept were revised in due course. Most notably, Varela and Coutinho \cite{Coutinho89,VC91,Coutinho03} suggested second generation networks with an architecture comprising a strongly connected central part with autonomous dynamics and a sparsely connected peripheral part for localized memory and adaptive immune response. In a sense, they reconcile both paradigms of idiotypic networks and clonal selection. A readable history of immunological paradigms can be found in \cite{Carneiro97}. Reviews with focus on idiotypic networks are \cite{Behn07} with an emphasis on modeling approaches and more recently \cite{Behn11} with emphasis on new immunological and clinical developments.

Today the main activities are in clinical research. Idiotypic interactions are the base of all therapies with mono\-clonal antibodies \cite{Reichert11}. New experimental techniques make large-scale studies of the idiotypic repertoire feasible \cite{Madietal11} which are necessary to infer the networks architecture.


In this paper we consider a minimalistic model of the idiotypic network, which was first formulated and investigated in \cite{BB03}. In this model a node represents lymphocytes and antibodies of a given idiotype. Lymphocytes of complementary idiotype can stimulate each other. The corresponding nodes are connected by links. 

Idiotypes are represented by bitstrings \cite{[{Representation of idiotypes by bitstrings was proposed by }] [{. See however also }]  FPP86,*Jerne85} of length $d$, $\bm{b}_d \bm{b}_{d-1} \cdots \bm{b}_1$ with $\bm{b}_i \in \{0, 1\}$. Ideally, $d$ is chosen such that $2^d$ is the size of the potential repertoire. The nodes of the network are labeled by these bitstrings. Two nodes $v$ and $w$ are linked if their bitstrings are complementary allowing for up to $m$ mismatches, i.e. their Hamming distance is $d_H (v, w) \geqslant d-m$. The corresponding undirected graph is the base graph $G_d^{(m)}$. Each node has the same number of neighbors, $\kappa = \sum_{k=0}^m {d \choose k}$. It represents the potential idiotypic repertoire with all possible interactions.

Not all idiotypes are expressed in the real network. In our minimalistic model we only account whether an idiotype is present or not, correspondingly the node is occupied, $n(v)=1$, or empty, $n(v)=0$. The subgraph of $G_d^{(m)}$ induced by the occupied nodes represents the expressed idiotypic network at a given time.

We describe the temporal evolution in discrete time. The influx of new idiotypes from the bone marrow is modeled by occupying empty nodes with probability $p$
\footnote{In the original version \cite{BB03} a constant number $I$ of new idiotypes was added in each influx step. Here we prefer to occupy empty nodes with a given probability $p$. This roughly corresponds to the transition from a microcanonical to a canonical approach in statistical physics.}.
For survival a B-lymphocyte needs stimulation by complementary structures. The dose-response curve is log-bell shaped, cf. \cite{PW97} and Refs. therein. If there are too many complementary structures, crosslinking becomes less likely due to steric hindrance, the stimulus is reduced. In our model an occupied node survives if the number of occupied neighbors is between two thresholds, $t_L$ and $t_U$. The rules of parallel update are
\begin{enumerate}
\vspace{-1mm}
\item[(i)] Occupy empty nodes with probability $p$\\[-6.5mm]
\item[(ii)] Count the number of occupied neighbors $n(\partial v)$ of node $v$. If $n(\partial v)$ is outside the window $[t_L, t_U]\,$, set the node $v$ empty\\[-6.5mm]
\item[(iii)] Iterate.\\[-5.5mm]
\end{enumerate}



Driven by the random influx of new idiotypes the network evolves towards a quasi-stationary state of nontrivial, functional architecture in which groups of nodes can be identified according to their statistical properties. Crucial for that is besides the random occupation of empty nodes, that occupied nodes are emptied if linked with too few or too many occupied nodes.


The paper is organized as follows. In the next section we discuss the model in its scientific context in other disciplines and its relation to other models of idiotypic networks. In Sec.~\ref{sec:simulationresults} we provide simulation results. We sketch a typical random evolution of the system to make the reader familiar with the systems behavior. Considering global and local network characteristics the existence of groups of nodes which share statistical properties \cite{BB03} is confirmed and more details are revealed.
In Sec.~\ref{sec:patternmodules} we describe certain regularities in the bitstrings of nodes which belong to the same group. The observed patterns can be constructed from pattern modules \cite{[{Preliminary accounts were given in }] SB06,*SB08}. The construction principle is explained first for the simplest pattern and generalized afterwards. These findings allow to calculate the number of groups, the group sizes, and the linking between groups. A new observable, the center of mass, is introduced that proves very useful in real-time pattern identification.
In Sec.~\ref{sec:specificpatterns} we apply this concept considering specific patterns observed in simulations, among them a dynamic pattern with core groups, peripheral groups, stable holes, and singletons that resembles in some aspects the biological network \cite{Coutinho89,VC91}.
In the appendix we calculate the scaling of the relative size of these groups for systems of biological size.

\section{Context and Related Models\label{sec:context}}

The topic falls into several scientific disciplines. 
It is natural to place our investigations in the context of network theory. Network theory has applications in a plethora of different, multidisciplinary fields \cite{Strogatz01,DM03,BA02} and has received great attention in the community of statistical physicists in the last decade.

A major body of research deals with growing
networks, where new nodes are attached to the existing nodes
randomly, or depending on properties of
the existing nodes. In this context
deletion of nodes is only
considered to study the resilience of the network against random or targeted attacks \cite{AJB00}. For recent reviews see \cite{dSS05,BLMCH06,Newman10}, cf. also \cite{DM03,Caldarelli07}. 

Natural networks however do not grow without limit but
stay finite and evolve towards a functional architecture. 
There are several modes to enable evolution:
(i)~Adding nodes but keeping the growth balanced by
deletion \cite{DT04,SMB05,SHL06,MGN06,BK07,GS08} or merging of
nodes \cite{KTMS05,AD05,SBM06}. (ii)~Keeping the nodes
unaffected but add, delete, or reorganize the links
\cite{DM00,WDeW04,ZKWDeW06,PLY05,LJ06,HNA08}. (iii)~Addition and
deletion of both nodes and links
\cite{DM01,SP-SSK02,DEB02,KE02,SR04,CFV04,SLZZ06}. Again, this
can be done randomly or depending on the properties of the
nodes and its neighbors.

Generic observables to characterize networks include the degree distribution, centrality, betweenness, cliquishness, modularity, clustering coefficients, and diameter. For instance,  growing networks using preferential attachment have a power law degree distribution like many real world networks. However, the characteristic exponent of preferential attachment networks is larger than the one found in natural networks. This was a major motivation to study evolving networks.

In many natural networks nodes have individual properties which control their potential linking. Clearly, in our case this is the idiotype. Also protein networks, transcription networks and generally signaling networks belong to this class.

Nodes may have an internal state which can change depending on their neighborhood in the network, or on external influences.
This dynamics has a typical time scale that is shorter than the time scale for evolution of the network's architecture. The interplay of these processes came into the focus of research only in the last few years. For a recent review and a status report see \cite{GB08,GS09}.
Our model is a very early example where this interplay is studied \cite{BB03}. 
In the present paper we describe the building principles of the architecture.


Our network model is a Boolean network \cite{Kauffman93}, since each node can be only in one of two states, empty or occupied. The nodes are updated in consecutive time steps depending on its own and its neighborhood occupation.

The model is also a {\em cellular automaton}, for a comprehensive monograph see \cite{Ilachinsky01}.  More precisely, since the update depends only on the sum of the neighbor states and the state of the node itself, it is a {\em totalistic} cellular automaton.
Cellular automata naturally involve unoccupied nodes. In our model unoccupied nodes, holes, play an important role. We distinguish holes that could be occupied from stable holes which can not be occupied due to overstimulation. A simple network picture disregards nodes which are not occupied.

A famous example of a totalistic two-dimensional deterministic cellular automaton is Conway's Game of Life \cite{Gardner70}. Both, the survival of an occupied cell and the occupation of an empty cell, are governed by window rules. Many interesting patterns, depending on the initial conditions static or dynamic, have been described in detail and classified.
The Game of Life was transfered to a variety of lattices, e.g. to three dimensional cubic lattices \cite{Bays87, *Bays06, Bays10ALT}, to triangular, pentagonal and hexagonal tesselations, to Penrose tilings \cite{Bays10ALT, Bays94, *Bays05, OS10ALT} and to small world geometry \cite{HZTJ03}. Larger than Life \cite{Griffeath94, Evans96, Evans01} increases the radius of the neighborhood.
Also, a probabilistic version on a 2d square lattice has been proposed, where stochastic deviations from the deterministic update are permitted \cite{SS78}. 
The mean occupation of cells, a global order parameter, undergoes a sharp phase transition for increasing strength of stochasticity. On the occasion of the 40th anniversary appeared a comprehensive collection of recent results on the Game of Life and its descendants \cite{Adamatzky10}.
Our model can be considered as a further version of the Game of Life on a high dimensional graph, where empty nodes are randomly occupied, while the survival of occupied nodes is governed by a deterministic window rule. Starting from an empty graph we observe an evolution toward a complex, highly organized architecture.

The model can also be categorized as a stochastic, non-linear dynamical system. 


There exists a variety of models for B-cell networks. References \cite{PW97,Behn07} give comprehensive surveys of modelling approaches. 

For instance, Stewart and Varela \cite{SV91,DBSV94} proposed a model, which also has a random influx and a window update rule to simulate the internal dynamics and a zero/one clone population. However, while we consider a discrete $d$-dimensional hypercubic shape space, in their model the complementary idiotypes live on different sheets of a 2D continuous shape space \footnote{In \cite{SV91} regular arrays of idiotype-anti-idiotype populations are observed, reminding in a sense of our 2-cluster pattern.}.
A summary of results obtained in models with continuous shape space is given by Bersini \cite{Bersini03}. Several aspects of modeling in continuous and discrete shape space are discussed in \cite{Hart06}.

An early network model inspired by spin glass physics was proposed by Parisi \cite{Parisi90} to describe immunological memory. The interaction between idiotypes in the model of Barra and collaborators \cite{BA10a,BA10b,BFS10} is also taken from spin glass physics.
Their model describes a given number of idiotype populations each with a constant number of lymphocytes. Each cell can be in a firing or quiescent state.  The strength of the ferromagnetic coupling between the idiotypes (also encoded by bitstrings) models the affinity, which is related to the complementarity of the bitstrings. Barra and Agliari \cite{BA10a,BA10b} compute the degree distribution, the type and number of loops, and consider the scaling behavior. They describe primary and secondary immune responses and understand low and high dose tolerance as a network phenomenon.
In our model the log-bell-shaped response is integrated as a B-cell property \cite{PW97,VC91}. Essential for our approach is a random influx of new cells and a selection mechanism. Hence, we can describe the evolution of the network towards a functional architecture.

A model which distinguishes between antibody molecules and lymphocytes with a dynamics including an influx similar to ours is proposed by Ribeiro et al. \cite{RDBV07}. They compare the dynamics on given architectures of random and scale-free networks.

IMMSIM, invented by Celada and Seiden \cite{CS86, PKSC02}, is also a modified cellular automaton. It incorporates many immunologic agents, including antigen presenting cells, B-cells, T-cells, antigens, antibodies and cytokines. It describes both humoral and cellular responses. Idiotypes are also characterized by bitstrings. The vast amount of interacting agents increases drastically the number of parameters. The model is intended to be as realistic as possible and to provide experimental and practical immunologists with a tool to test hypotheses {\em in silico}. 
Our approach is in a sense complementary. We aim at an understanding of the principles governing the autonomous evolution toward a functional architecture. Therefore we investigate a minimalistic model with a small number of parameters, which nevertheless exhibits essential features of the biological network.

The concept of idiotypic networks inspired applications in computer sciences, e.g. artificial immune systems for the detection of intrusion, of spam or viruses \cite{FB07,ICARIS10}.

\section{Simulation Results\label{sec:simulationresults}}

In this section we report on simulations on the base graph $G_{12}^{(2)}$, which consists of $2^{12}=4096$ nodes, each of which has $\kappa = 79$ links to other nodes. The window rule parameters are $[t_L,t_U]=[1,10]\,$. The lower threshold is biologically motivated, a node needs at least one occupied neighbor to survive. The upper threshold is chosen to enable a  non-trivial, dynamic pattern of complex architecture. As discussed below in more detail, higher $t_U$ would allow a broader variety of static patterns. However, the concept developed in this paper still applies.

The simulations start with an empty base graph. The influx $p$ varies from $0$ to $0.11$. This covers the range in which we find interesting patterns, above $p\gtrapprox t_U/\kappa = 0.127$ there is only trivial random behavior.

\subsection{Random Evolution\label{sec:randomevolution}}

We start the simulations with an empty base
graph occupying nodes with probability $p$.  In the first time
step only those nodes survive which have at least one occupied
neighbor (having more than $t_U$ occupied neighbors is unlikely in the
beginning).  The surviving nodes represent seeds the neighbors of
which will survive if occupied.  Hence, we observe a rapid growth of a
giant cluster until more and more nodes have more than $t_U$ occupied
neighbors.  Exceeding the upper threshold deletes a node.  Thus, the
giant cluster decays and many stable holes are created, i.e.  nodes
with the number of occupied neighbors above the upper threshold $t_U$.
Fig.~\ref{fig:timeseries} shows a time series for the evolution to a stationary state where the number of stable holes increases up to its stationary value.

The empty base graph is a highly symmetric object. Due to the random
influx the symmetry is broken and the system falls into a network
configuration of lower symmetry. 
The typical result of the evolution is a quasi-stationary pattern of mutually dependent occupied nodes and stable holes. Properly situated occupied nodes create stable holes and in return stable holes in the neighborhood of an occupied node may prevent its overstimulation.

In the generic case for a given $p$ a certain pattern type is found most 
frequently. Occasionally, depending on the history of the driving process,
also other patterns occur. Once established they all can live for a long time. (This can be proved in simulations preparing the pattern as initial configuration.)

In principle, the system is ergodic \footnote{In this property the original version of the model \cite{BB03} with a constant number of new idiotypes in each influx step can be different, obviously.}. This becomes immediately clear considering the following unlikely but possible event. For any choice of $p>0$ the occupation of nodes by the influx can be such that the application of the window rule leads to an empty base graph. After this extinction catastrophe the further realization of the driving process, the influx, determines to which pattern the system evolves. For any $p>0$ an infinite trajectory contains partial trajectories leading to any possible pattern. In practice, in our simulations we have never seen such an extinction catastrophe but only quasi-stationary patterns usually living for a long time. For increasing system size such catastrophes become less likely. In the thermodynamic limit we expect a breaking of ergodicity.

In our finite system there can be transitions between different patterns on a route without extinction catastrophe but via the formation of an intermediate, unstable giant cluster, for more details see \cite{BB03}.

Modeling biological systems we should keep in mind that they are finite and have a finite life expectation. The quasi-stationary state could persist for times longer than the life span of the individual but transitions between different states can not be excluded. Moreover, the parameters could vary during the individual's life.

\subsection {Global Characteristics\label{sec:globalchar}}

A first characterization of the different patterns can be obtained considering global quantities. They include the number of occupied nodes on the base graph 
\begin{equation}
n(G) = \sum _{v\in G} n(v), 
\end{equation}
where $n(v) \in \{0,1\}$ is the occupation of node $v$, the size of the
largest cluster in the set of present clusters $\cal C$
\begin{equation}
|C^{\max}| = \max_{C\in \cal C} (|C|),
\end{equation}
and the average size of the clusters 
\begin{equation}
\langle |C| \rangle_{\cal C} = \frac {1}{N_C} \sum_{C \in \cal C} |C|, 
\end{equation}
where $N_C$ is the current number of clusters, and $\langle \;\;\cdot \;\; \rangle _{\cal S}$ denotes the average over the elements of some set ${\cal S}$. Clusters, i.e. connected parts of the occupied subgraph, are very characteristic for patterns. Finally, we mention the
number of stable holes $h^*(G)$, i.e. empty nodes with $n(\partial v)>t_U$.

\begin{figure}[t]
\centering 
\includegraphics[height=\columnwidth, angle=270]{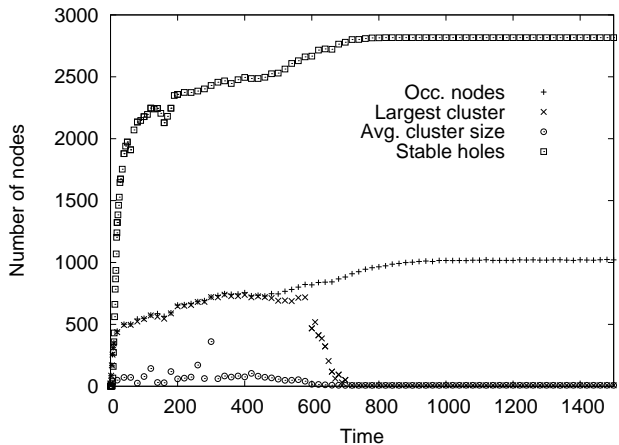}
\caption{Time series of the number of occupied nodes $n(G)$, the size of the currently largest cluster $|C^{\max}|$, the average cluster size $\langle |C| \rangle_{{\cal C}}$, and the number of stable holes $h^*(G)$ on the base graph $G_{12}^{(2)}$ with $[t_L,t_U] = [1,10]$ and $p\!=\!0.01$. The system evolves to a stationary 8-cluster pattern.}
\label{fig:timeseries}
\end{figure}

Figure~\ref{fig:timeseries} shows a generic time series of these global characteristics for a parameter setting where the system evolves to a stationary 8-cluster pattern. 

In the stationary state we can consider temporally averaged global quantities. The temporal average is defined as 
\begin{equation}
\overline{x} = \frac{1}{T_1-T_0} \sum_{t\in (T_0,T_1]} x_t, 
\end{equation}
where $T_0$ should be larger than the relaxation time in which the system reaches the
stationary state, and $(T_0, T_1]$ is the averaging period.

Table~\ref{tab:globalq} gives results for three patterns which occur for different influx $p$. The static patterns are named according to the size of the characteristic clusters. With a deeper understanding of the architecture we shall name them by the number of groups of nodes, see Sec.~\ref{sec:patternmodules} below.

\begin{table}[th]
\caption{\label{tab:globalq} Temporal averages and standard deviations of global characteristics for three typical patterns. Data from 500,000 iterations.\\}
\begin{ruledtabular}
\begin{tabular}{c*{3}{@{\quad}r@{$\,\pm\,$}l}}
  Pattern & \multicolumn{2}{c}{8-cluster} &
  \multicolumn{2}{c}{2-cluster} & \multicolumn{2}{c}{dynamic} \\ 
  $p$ & \multicolumn{2}{c}{0.005} & \multicolumn{2}{c}{0.015} &
  \multicolumn{2}{c}{0.025} \\ 
  \hline
  $\overline{n(G)}$ & 1024.6 & 0.8 & 1023.3 & 0.9 & 533.4 & 27.7 \\
  $\overline{h^*(G)}$ & 2816.0 & 0.0 & 3055 & 20 & 1828
  & 114 \\
  $\overline{|C^{\max}|}$ & 8.0 & 2.8 & 4.5 & 2.1 & 410 & 20 \\
  $ \overline{\langle |C| \rangle}_{\cal C}$ & 7.96 & 0.04 & 2.01
  & 0.01 & 4.49 & 0.96 \\
  $\overline{N}_C$ & 128.6 & 0.8 & 510.4 & 3.7 & 123.1
  & 22.0 \\
\end{tabular}
\end{ruledtabular}
\end{table}

The 8-cluster pattern at $p\!=\!0.005$ has 128 clusters of size 8,
which results in a total average population of 1024 occupied
nodes, i.e. one fourth of all nodes.
The remaining nodes are holes, 2816 of which are stable ($n(\partial
v)\!>\!t_U$), and 256 are not stable ($n(\partial v)\!\leqslant\!t_U$). Empty nodes
with $n(\partial v)\!<\!t_L$ are not counted as stable
holes, since they could easily become occupied and sustained by the
random influx. Figure~\ref{fig:timeseries} shows the temporal
evolution from an empty base graph towards an 8-cluster pattern.

\begin{figure}[t]
  \centering 
  \includegraphics[width=0.8\columnwidth]{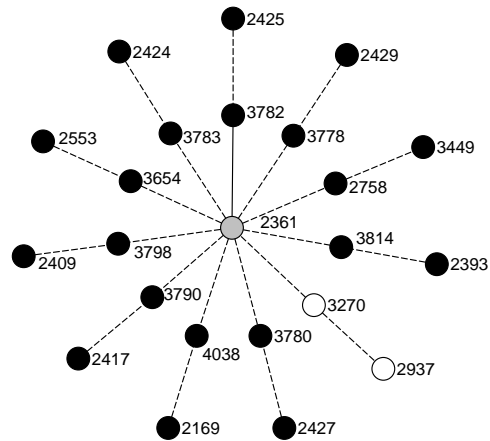}
  \caption{A selection of 2-clusters as they appear for moderate influx. We see ten 2-clusters of occupied nodes (\textit{black}). They are all connected through a hub (\textit{gray}). If the eleventh 2-cluster (\textit{empty circles}) becomes occupied, the hub would not survive the next update. There is one link (\textit{solid}) with one mismatch, the other links (\textit{dashed}) have two mismatches. The nodes are labeled with the decimal expressions of their bitstrings. Figure produced using yEd \cite{yWorks}.}
\label{fig:gamma2c}
\end{figure}

In the 2-cluster pattern at $p\!=\!0.015$ the 510 clusters of size 2
together occupy about one fourth of the base graph. The remaining three
quarters are stable holes. Defects in the perfect 2-cluster pattern allow that occasionally some of the node pairs become connected via a central hub and a larger star-shaped
cluster is formed, see Fig.~\ref{fig:gamma2c}. $t_U=10$ was chosen to exclude the occupation of the hub in the perfect pattern.
Both the 8- and the 2-cluster patterns are quasi-static, the temporal fluctuations are small.

\begin{figure}[t]
\centering
\includegraphics[width=\columnwidth]{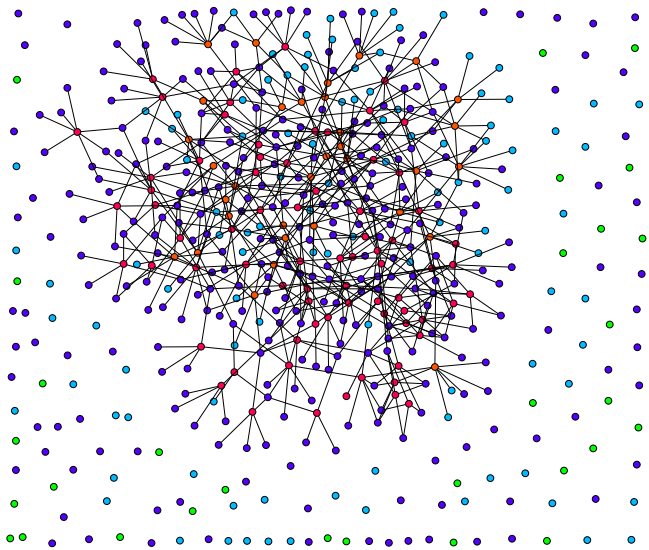}
\caption{ (Color online) Snapshot of the occupied graph of the complex configuration of a dynamic pattern for $p=0.025$. The nodes are colored according to their membership in different groups, see Fig.~\ref{fig:scheme12gb} (top). Nodes within a group have similar statistical characteristics, see text. Figure produced using yEd \cite{yWorks}.}
\label{fig:gamma6g}
\end{figure}

A more complex, dynamic pattern evolves for larger $p\gtrapprox0.03$. The
standard deviation of the temporal averages is one order of magnitude larger than in the
two static patterns. Figure~\ref{fig:gamma6g} shows a snapshot of the occupied graph. We see one large cluster of about 400 nodes and about 120 isolated nodes. These nodes had at least one occupied neighbor which was removed in latest update. Further, approximately 1800
stable holes and 1800 unstable holes have been observed, which are not shown in the figure. We can clearly distinguish a central and a peripheral part, which are supposed to be
of functional importance in the biological idiotypic network. 

\subsection{Local Characteristics\label{sec:localchar}}

Besides global quantities, time averages of local quantities characterizing every single node can be considered. Elucidating are the mean occupation $\overline{n}(v)$, the number of occupied neighbors $\overline{n}(\partial v)$, and the mean life time $\overline{\tau}(v)$, which is defined as
\begin{equation}
  \overline{\tau}(v) = \frac{1}{b(v)+n_{T_0}} \sum_{t\in (T_0, T_1]} n_t(v)\,,
\end{equation}
where $b(v)$ is the number of births during the observation time,
i.e. the number of new occupations of the node by the influx. Of
course, $b(v)+n_{T_0} \neq 0$ must be fulfilled, otherwise
$\overline{\tau}(v)$ has no meaning.

\begin{figure}[t]
\centering
\includegraphics[height=\columnwidth, angle=270]{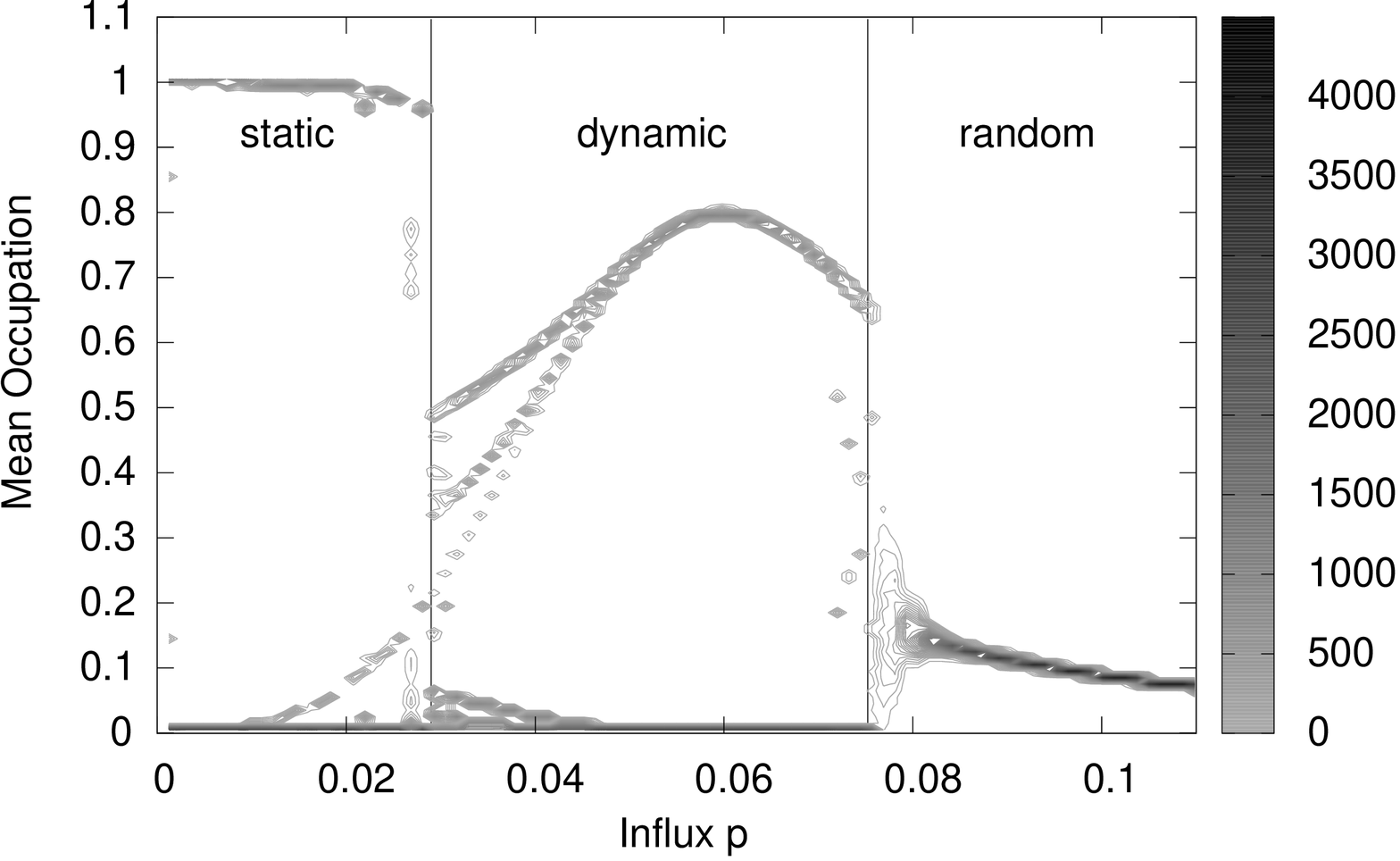}
\includegraphics[height=\columnwidth, angle=270]{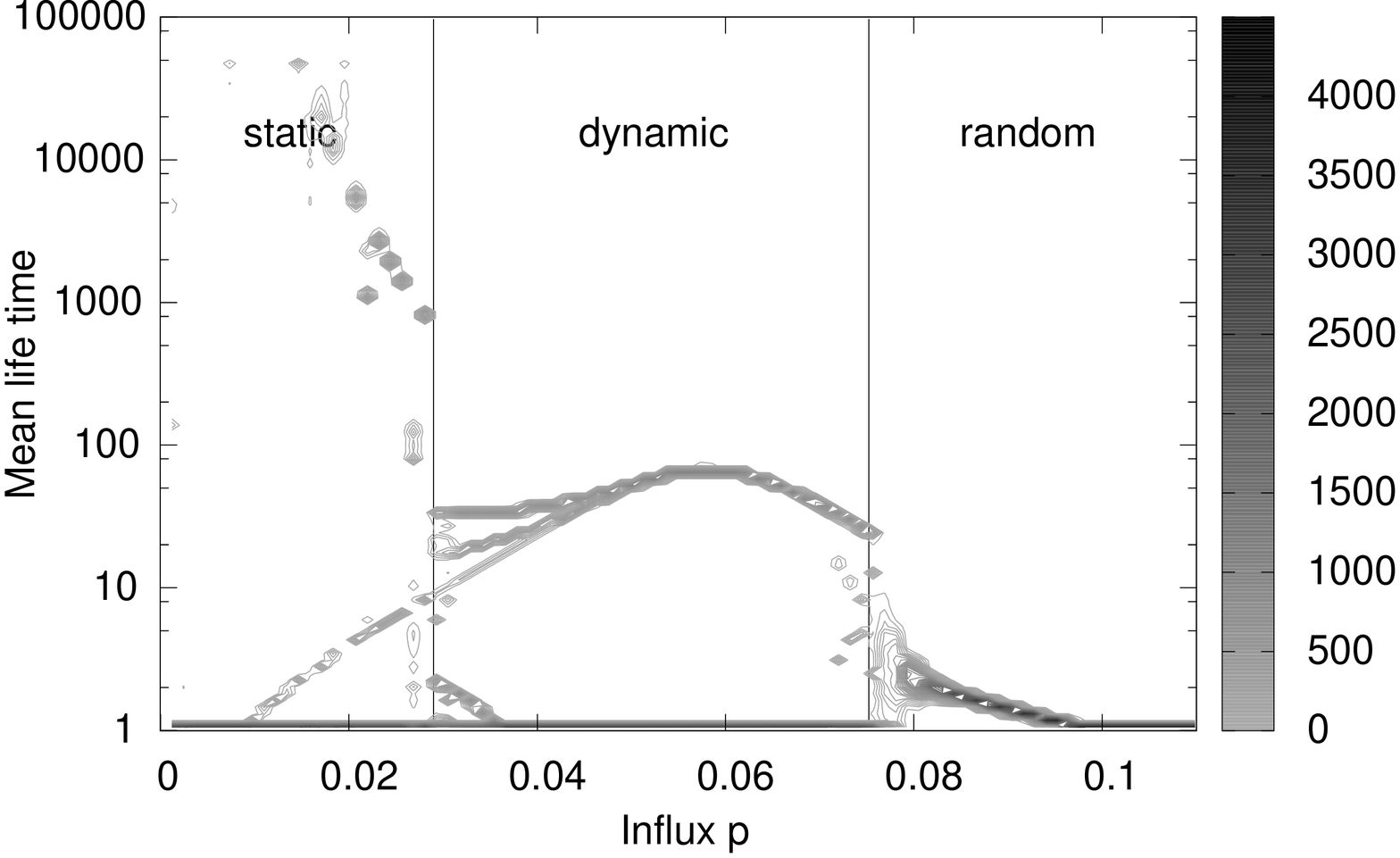}
\includegraphics[height=\columnwidth, angle=270]{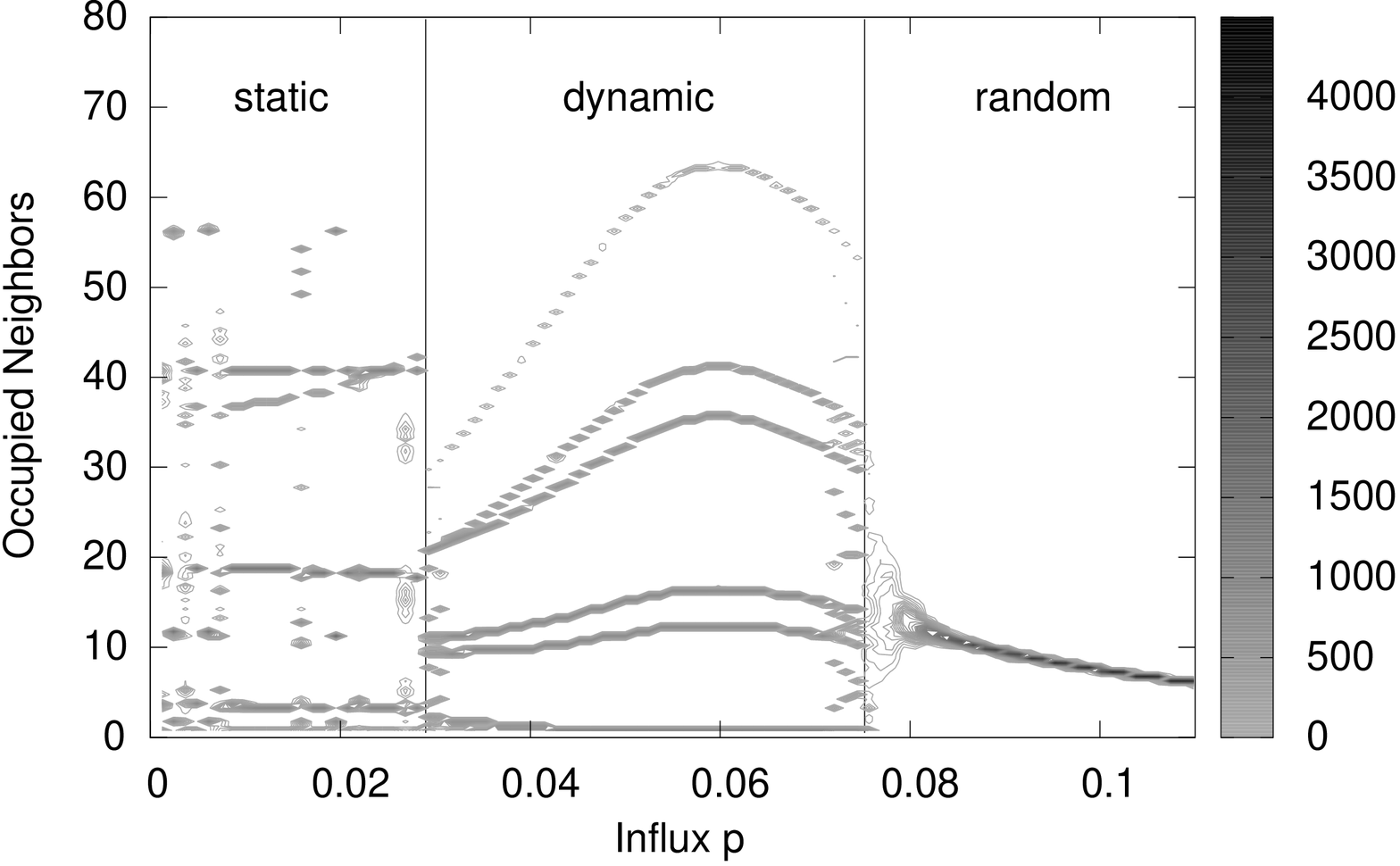}
\caption{Time averages of the mean occupation (top), the mean life time (center), and the mean number of occupied neighbors (bottom) of each node for increasing values of $p$, $\Delta p = 5/4096$. Each graph shows isolines of a histogram in top view. The grey scale reflects the number of nodes with the corresponding property. Regimes of different temporal behavior are separated by vertical lines. For each $p$ data is from time series of 500,000 iterations in the stationary state, starting from an empty base graph $G_{12}^{(2)}$.}
\label{fig:hist} 
\end{figure}

We can identify groups of nodes sharing statistical properties as proposed in \cite{BB03}. Figure~\ref{fig:hist} shows mean occupation, mean life time, and the number of occupied neighbors vs. influx probability. The groups appear as peaks in the histograms. The number of occupied neighbors proves most suitable to distinguish the different groups.

For small and moderate influx a clear group structure is visible.  The data for the mean life time and the mean occupation suggest that the patterns are static for $p \lessapprox 0.03$. 
These patterns have groups of occupied nodes with a high mean life time, other groups are stable holes or sparsely occupied nodes. For $0.03 \lessapprox p \lessapprox 0.08$ the patterns are dynamic, but still stationary. Also in dynamic patterns there are stable holes. The mean life time of occupied nodes is small, the occupied subgraph changes continuously. 
While in static and dynamic patterns all nodes remain in their groups, for high influx $p \gtrapprox 0.08$ the patterns become short-lived, groups dissolve and reemerge in a different configuration. Their distinction by temporal averages becomes more and more difficult. For very high influx $p \gg 0.12$ the dynamics is entirely random.

\section{Concept of Pattern Modules\label{sec:patternmodules}}

We find regularities in the bitstrings encoding nodes belonging to the same group. This allows to identify general building principles. Patterns are build from more elementary objects, so called {\em pattern modules}. With this concept we can derive all structural properties, the size of the groups and their linking, of the patterns observed so far in the simulations.

We first explain the principles on the example of the 2-cluster pattern, and develop then the detailed concept for the general case. At the end of the section, exploiting this concept we introduce the center of mass vector, a tool which allows to identify in simulations the majority of patterns in real time.

\subsection{Simplest Case: The 2-Cluster Pattern
  \label{sec:2cluster}}

We introduce the building principles of patterns considering the simplest pattern, the 2-cluster pattern which appears for moderate influx. In simulations starting from empty base graphs it is rare and only found for $0 < p \lessapprox 0.03$, but if prepared as initial condition it is very stable for a larger range of $p$ up to $0.045$. By their statistical characteristics, cf. Table~\ref{tab:groups2c}, we can distinguish three groups of nodes, frequently {\em occupied nodes} ($S_1$) with a high mean life time, permanently empty {\em stable holes} ($S_3$), and rarely occupied {\em potential hubs} ($S_2$) which link together up to $t_U$ 2-clusters if occupied.

\begin{table}[h]
\caption{Local characteristics of the three groups in the 2-cluster pattern 
for $p\!=\!0.025$ compared with the ideal pattern. Data from a time series of 500,000 iterations.\\}\label{tab:groups2c}
\begin{ruledtabular}
\begin{tabular}{lc@{\qquad}c@{\quad}c@{\quad}c}
& & $S_1$ & $S_2$ & $S_3$ \\[1mm]
\hline\\[-2mm]
mean occupation & $\langle \overline{n}(v) \rangle_{S_i}$ &
0.993 & 0.0004 & 0.000 \\
& $n_{\text{ideal}}$ & 1 & 0 & 0 \\[1mm]
\hline\\[-2mm]
occ. neighbors & $\langle \overline{n}(\partial v) \rangle_{S_i}$ &
1.002 & 10.95 & 55.64 \\
& $n(\partial v)_{\text{ideal}}$ &
1 & 11 & 56 \\[1mm]
\hline\\[-2mm]
mean life time & $\langle \overline{\tau} (v) \rangle_{S_i}$ &
6923 & 0.016 & 0.000 \\[1mm]
\end{tabular}
\end{ruledtabular}
\end{table}

Looking at the node indices $i_v\,$ in decimal representation we
observed that the sum of the two indices in a 2-cluster is constant in a
realization. In a different realization the index sum can be different.
For instance, in Fig.~\ref{fig:gamma2c} the index sum within all 2-clusters is
6207. This indicates regularities at the level of the bitstrings.

We found that all occupied nodes are identical in exactly two bits,
say at position $k$ and $l\,$.  The members of a 2-cluster are
complementary in all other bits, in symbols we write
\begin{equation} \label{eq:2connect} \bm{\cdots b}_k \bm{\cdots b}_l \bm{\cdots}
  \quad \text{is linked with} \quad \bm{\overline{\cdots} b}_k
   \bm{ \overline{\cdots} b}_l \bm{\overline{\cdots}} \;, \nonumber
\end{equation}
where the bar denotes the bit inversion. The bitstrings of all stable holes are
also equal in the same two bit positions $k$ and $l\,$. However, they
are inverse to $\bm{b}_k$ and $\bm{b}_l$ of the occupied nodes.
Potential hubs have exactly one inverse and one equal bit in these
positions. As these bits play a crucial role, we call them {\em determinant bits}. 
The regularities are summarized by
\begin{displaymath}
\begin{array}[b]{r@{\quad}l}
\text{occupied nodes}\ S_1 & \quad\,
\bm{\cdots} \bm{b}_k \bm{\cdots} \bm{b}_l \bm{\cdots} \\
\text{potential hubs}\ S_2 & \left\{
\begin{array}{c} 
\bm{\cdots} \overline{\bm{b}_k} \bm{\cdots} \bm{b}_l \bm{\cdots} \\
\bm{\cdots} \bm{b}_k \bm{\cdots} \overline{\bm{b}_l} \bm{\cdots}
\end{array} \right. \\
\text{stable holes}\ S_3 & \quad\,
\bm{\cdots} \overline{\bm{b}_k} \bm{\cdots} \overline{\bm{b}_l}
\bm{\cdots} 
\end{array} \; .
\end{displaymath}
The example in Fig.~\ref{fig:gamma2c} has the determinant bits
in positions 7 and 12, $\bm{b}_7=\bm{b}_{12}=1$.

This allows to explain all structural properties of the pattern observed in the simulations. We can construct an ideal 2-cluster pattern, a configuration in which all nodes of group $S_1$
are occupied and the others remain empty. It is ideal in the sense
that there are no defects but also no hubs.

Since all other bits can take all possible combinations, the size of the groups can be calculated, e.g. there are $|S_1|=2^{d-2}$ occupied nodes and $|S_2|=2\times 2^{d-2}$ potential hubs.

\begin{figure}[t]
\centering
\includegraphics[width=0.7\columnwidth]{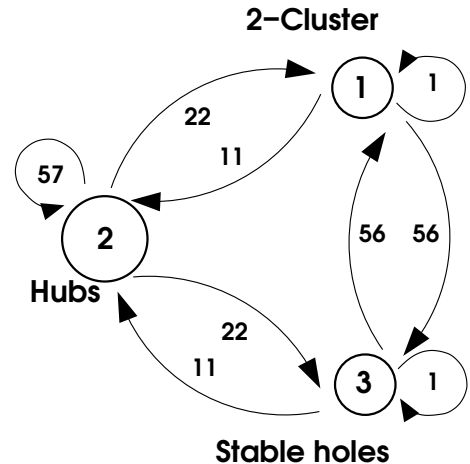}
\caption{The three groups of the 2-cluster pattern and their linking. The circle sizes
  correspond to the group sizes. The arrows and the numbers next to
  them indicate how many nodes of a group exert influence on the nodes
  of another group, e.g. each node of $S_2$ is stimulated by 11 nodes from $S_1$. The 
  number of links can be counted in simulations or taken from the link matrix that is
  derived in Sec.~\ref{sec:generalcase}.}
\label{fig:scheme3g}
\end{figure}

We further can compute the number of occupied neighbors $n(\partial
v)$ of a node $v$ of any group. Since all nodes of $S_1$ are
occupied in the ideal pattern, $n(\partial v)$ is given by the
number of links between $v$ and nodes in $S_1$.  A link
between two nodes exists if their bitstrings are complementary except
for up to two mismatches.  If $v \in S_1\,$, it has two bits in common
with all other nodes in $S_1$, namely $\bm{b}_k$ and
$\bm{b}_l\,$.  Thus, all remaining bits must be exactly
complementary. There is only {\em one} node $w \in S_1$,
which obeys this constraint.  If $v \in S_2$ or $v \in S_3$, there is
one pre-determined mismatch or none, respectively. The remaining
mismatches can be distributed among the $d-2$ non-determinant bits.
Thus a node in $S_i$ has
\begin{equation}
n(\partial v)_{\text{ideal}} = \sum_{j=0}^{i-1} {d-2 \choose j}
\end{equation}
occupied neighbors in the ideal pattern. For small influx this is in good agreement with the simulations, cf. Table \ref{tab:groups2c}. In a similar way  for all nodes the number of links to nodes in different groups can be calculated, the result is visualized in Fig.~\ref{fig:scheme3g}. The derivation for the general case is given in the next subsection. 

\begin{figure}[t]
\centering
\includegraphics[width=0.9\columnwidth]{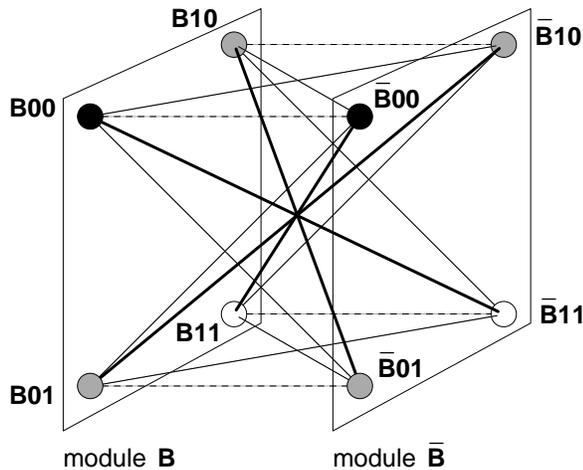}
\caption{Two pattern modules with complementary non-determinant bit chains $\bm{\mathsf{B}}$ and $\bm{\overline{\mathsf{B}}}$, respectively, on a two mismatch base graph with a 2-cluster configuration. The two-dimensional modules are congruently occupied, each consisting of one occupied node (\textit{black}, $\bm{\cdot 00}$), two potential hubs (\textit{gray}, $\bm{\cdot 01}$ and $\bm{\cdot 10}$) and one stable hole (\textit{white}, $\bm{\cdot 11}$). The positions and values of the determinant bits are chosen without loss of generality. The links have no mismatch (\textit{bold}), one (\textit{solid}), or two mismatches (\textit{dashed}). There are $2^{d-2}$ pattern modules, each pair of which with complementary non-determinant bits contribute a pair of occupied nodes.}
\label{fig:pmodules}
\end{figure}

This regularity encouraged the following concept. Considering the two
determinant bits as coordinates of a two-dimensional space, they
define the corners of a two-dimensional hypercube.  The corner with
coordinates $(\bm{b}_k, \bm{b}_l)$ represents an occupied node, the
opposite corner $(\overline{\bm{b}_k}, \overline{\bm{b}_l})$ is a
stable hole, and the neighboring corners of $(\bm{b}_k, \bm{b}_l)$ are
potential hubs. We call this structure a {\em pattern module}, because
it is the building block for the entire regular configuration. In a
different picture, we can also understand an ideal configuration as
consisting of $2^{d-2}$ congruently occupied `parallel worlds'. 
Figure~\ref{fig:pmodules} illustrates the concept of pattern modules.

Any choice of the two determining bits is of course possible, all
corresponding patterns are equivalent, the 2-cluster pattern is $2^2
\times {d \choose 2}$-fold degenerated, where the first factor
represents the choice of the two determinant bits, and the second
factor gives the number of possible positions of these bits in the
bitstring of length $d\,$.  It is the individual history (the
realization of the random influx), which selects the determining bits
and thus breaks the symmetry.

\subsection{General Case\label{sec:generalcase}}

\subsubsection{Groups\label{sec:groups}}

Many results for 2-cluster patterns on the $G_{12}^{(2)}$ base graph
can be generalized to more complex architectures and other choices of
$d$ and $m$. This includes the 8-cluster pattern mentioned in
Sec.~\ref{sec:simulationresults} and other static patterns as well as the dynamic
pattern. Their structure is correctly described by pattern modules
with more than two determinant bits.

In the same way as for the 2-cluster pattern, we define the pattern
module as a hypercube of dimension $d_M$, where $d_M$ is the number of
determinant bits. There
are two groups which are represented by only one node in the
pattern module. One of them is labelled $S_1$. 
All nodes in group $S_1$ have the same determinant bits $\bm{b}_1 \cdots
\bm{b}_{d_M}$. The other groups are ordered such that the determinant bits
of $S_j$ differ in $j\!-\!1$ positions from the determinant bits of
$S_1$. It is clear that there are $d_M\!+\!1$ groups. Groups $S_j$ and $S_{d_M+2-j}$
are equivalent, they have the same size and linking properties, see below.
In typical patterns, occupation breaks the symmetry. We usually label the 
smallest group of the more occupied half as $S_1$, which is of course arbitrary.

The size of the groups can be obtained by elementary
combinatorics.  Since, the $i$-th group deviates in $i\!-\!1$ out of $d_M$
determinant bit positions of $S_1$, there are $d_M \choose
i-1$ choices. This number is multiplied by the number of pattern modules on
the base graph given by $2^{d\!-\!d_M}$. The group size is
\begin{equation} \label{eq:groupsize}
  |S_i| = 2^{d-d_M} {d_M \choose i-1}\,,\ i=1, \dots , d_M\!+\!1\,.
\end{equation}
The factor ${d_M \choose i-1}$ is called the relative group size,
since it is the group size normalized by the size of $S_1$. It is
independent of the base graph dimension $d$ and the number of
mismatches $m$.

As an example, we can construct 2-cluster patterns on a base graph
$G_d^{(m)}$ by means of pattern modules with exactly one occupied
node ($S_1$).  The dimension of the pattern module $d_M$ then has to
equal the number of allowed mismatches $m$. The number of
qualitatively distinguishable groups is $m\!+\!1$, etc.  A 2-cluster
pattern can emerge if the lower threshold is $t_L \leqslant 1\,$ and
the upper threshold obeys $1 \leqslant t_U \leqslant d-m\,$. The
2-cluster pattern on 1-mismatch graphs described in \cite{BB03} is an
instance of such a pattern. However, in
the 1-mismatch case one half of all nodes are occupied, the other half
are stable holes.

\subsubsection{Linking\label{sec:linking}}

Each node on $G_d^{(m)}$ has $\kappa=\sum_{k=0}^m {d \choose k}$ links, $\kappa$ is constrained by the allowed number of mismatches $m$. We consider a pattern with $d_M$ determinant bits. Each node in group $S_i$ is linked to $L_{ij}$ neighbors in group $S_j$. The $L_{ij}$ are the entries of the link matrix $\mathbb{L}$. $\mathbb{L}$ defines the architecture of a pattern built of modules of dimension $d_M$. 

The dynamics of a node depends on the number of occupied neighbors. The mean occupation is a typical common property of nodes belonging to the same group. Thus, the knowledge of the group membership of the node's neighbors is of crucial importance to understand its statistical properties. 

Within the concept of pattern modules $\mathbb{L}$ can be derived combinatorially.
Recall that the determinant bitstring of a node in $S_l$ deviates in $l\!-\!1$ bits from the determinant bitstring of a node $v_1$ in $S_1$. In the following we consider a node $v^{(i)}$ chosen such that the $i-1$ bits inverse to the corresponding bits of $v_1$ are left-aligned, its non-determinant bits are denoted by $\bm{\mathsf{B}}$. This choice is without loss of generality, because all the arguments do not depend on the labeling of the bit position. From the nodes in $S_j$ we choose $v^{(j)}$ such that the $j-1$ bits inverse to the corresponding bits of $v_1$ are right-aligned and the non-determinant bits are $\bm{\overline{\mathsf{B}}}$. There is no other node in $S_j$ with less mismatches to $v^{(i)}$.

We have to distinguish whether or not the partial bitstrings of length $i-1$ and $j-1$ do overlap. These cases are discriminated by the value of $\Delta_{ij} = d_M\!-\!i\!-\!j\!+\!2$. There are three cases.\\
(i) $\Delta_{ij}\!=\!0$. All determinant bits of $v^{(i)}$ and
$v^{(j)}$ are complementary.\\
(ii) $\Delta_{ij}\!>\!0$. $v^{(i)}$ and $v^{(j)}$ share $\Delta_{ij}$
determinant bits with $v_1$. This case is illustrated below.
\begin{displaymath}
\label{eq:detbitscheme}
\begin{array}{c} v_1 \\[2.2mm] v^{(i)} \\[2.2mm] v^{(j)} \end{array} \
\underbrace{
\begin{array}{ccc} 
  \bm{b}_1 & \dots & \bm{b}_{i-1} \\[2mm] \hline
  \bm{b}_1 & \dots & \bm{b}_{i-1} \\[2.2mm] 
  \bm{b}_1 & \dots & \bm{b}_{i-1} 
\end{array}
}_{i-1\ \text{bits}}
\underbrace{
\begin{array}{ccc} 
  \bm{b}_i & \dots & \bm{b}_{d_M\!-\!j\!+\!1} \\[2.2mm]
  \bm{b}_i & \dots & \bm{b}_{d_M\!-\!j\!+\!1} \\[2.2mm] 
  \bm{b}_i & \dots & \bm{b}_{d_M\!-\!j\!+\!1} 
\end{array}
}_{\Delta_{ij} = d_M-i-j+2\ \text{bits}}
\underbrace{
\begin{array}{ccc} 
  \bm{b}_{d_M\!-\!j\!+\!2} & \dots & \bm{b}_{d_M} \\[2.2mm]
  \bm{b}_{d_M\!-\!j\!+\!2} & \dots & \bm{b}_{d_M} \\[2mm] \hline
  \bm{b}_{d_M\!-\!j\!+\!2} & \dots & \bm{b}_{d_M}
\end{array} 
}_{j-1\ \text{bits}}
\end{displaymath}
(iii) $\Delta_{ij}\!<\!0$. $v^{(i)}$ and $v^{(j)}$ share $|\Delta_{ij}|$
determinant bits which are inverse to the corresponding bits of $v_1$, see diagram below.
\begin{displaymath}
\begin{array}{c} v_1 \\[2.2mm] v^{(i)} \\[2.2mm] v^{(j)} \end{array} \
\underbrace{
\begin{array}{ccc} 
  \bm{b}_1 & \dots & \bm{b}_{d_M\!-\!j\!+\!1} \\[2mm] \hline
  \bm{b}_1 & \dots & \bm{b}_{d_M\!-\!j\!+\!1} \\[2.2mm] 
  \bm{b}_1 & \dots & \bm{b}_{d_M\!-\!j\!+\!1} 
\end{array}
\underbrace{
\begin{array}{ccc} 
  \bm{b}_{d_M\!-\!j\!+\!2} & \dots & \bm{b}_{i-1} \\[2mm] \hline
  \bm{b}_{d_M\!-\!j\!+\!2} & \dots & \bm{b}_{i-1} \\[2mm] \hline
  \bm{b}_{d_M\!-\!j\!+\!2} & \dots & \bm{b}_{i-1} 
\end{array}
}_{-\Delta_{ij}=i+j-d_M-2\ \text{bits}}
}_{i-1\ \text{bits}}
\underbrace{
\begin{array}{ccc} 
  \bm{b}_i & \dots & \bm{b}_{d_M} \\[2.2mm]
  \bm{b}_i & \dots & \bm{b}_{d_M} \\[2mm] \hline
  \bm{b}_i & \dots & \bm{b}_{d_M}
\end{array} 
}_{d_M-i+1\ \text{bits}}
\end{displaymath}

The number of mismatches between $v^{(i)}$ and $v^{(j)}$ is $|\Delta_{ij}|$. If $|\Delta_{ij}| \leqslant m$ there is a link between $v^{(i)}$ and $v^{(j)}$.
In this case there are further nodes in $S_j$ which link to $v^{(i)}$, the number of which is calculated combinatorially. There are $m\!-\!|\Delta_{ij}|$ additionally allowed mismatches which can appear among non-determinant and/or determinant bits. 

In cases (ii) and (iii) there are further nodes in $S_j$ with $|\Delta_{ij}|$ mismatches to $v^{(i)}$. These are obtained by distributing the $|\Delta_{ij}|$ mismatches among the $d_M\!-\!i\!+\!1$ right-aligned bits in case (ii), or among the $i-1$ left-aligned bits in case (iii). This leads to ${d_M\!-\!i\!+\!1 \choose \Delta_{ij}}$ and ${i-1 \choose |\Delta_{ij}|}$ nodes in the respective cases.

Now we consider additional mismatches. Among the $d-d_M$ non-determinant bits we can distribute $l$ mismatches in $${d\!-\!d_M \choose l}$$ different ways.

Among determinant bits additional mismatches can only appear in pairs. This is relevant for $m\!-\!|\Delta_{ij}| \geqslant 2$. If we invert one bit in $v^{(j)}$, the resulting node is not in $S_j$ since the number of bits complementary to $v_1$ is changed. We have to invert a second bit so that the number of mismatches with $v_1$ remains constant. The number of nodes in $S_j$ with $|\Delta_{ij}|\!+\!2k$ mismatches to $v^{(i)}$ is computed as follows. 

In case (ii) we invert $k$ bits in the $i-1$ left-aligned bits of $v^{(j)}$. There are ${i-1 \choose k}$ possibilities. To compensate this, we have to invert $k$ bits among the $j-1$ right-aligned bits of $v^{(j)}$. Altogether, there are now $\Delta_{ij}\!+\!k$ mismatches to $v^{(i)}$ living on the $d_M\!-\!i\!+\!1$ right-aligned bits. There are ${d_M\!-\!i\!+\!1 \choose \Delta_{ij}\!+\!k}$ possibilities to distribute them. Thus we have $${i\!-\!1 \choose k}{d_M\!-\!i\!+\!1\choose \Delta_{ij}+k}$$ nodes in $S_j$ with $\Delta_{ij}+2k$ mismatches to $v^{(i)}$. 

Case (iii) is similar. We invert $k$ bits in the $d_M-i+1$ right-aligned bits of $v^{(j)}$, there are ${d_M-i+1 \choose k}$ possibilities.  To compensate this, we have to invert $k$ bits among the $d_M-j+1$ left-aligned bits of $v^{(j)}$. Together, $\Delta_{ij}+k$ mismatches to $v^{(i)}$ live on the $i-1$ left-aligned bits. There are ${i-1\choose |\Delta_{ij}|+k}$ possibilities. This leads to $${d_M\!-\!i\!+\!1 \choose k}{i-1\choose |\Delta_{ij}|\!+\!k}$$ nodes in $S_j$ with $|\Delta_{ij}|+2k$ mismatches to $v^{(i)}$.

The result in case (i) is obtained setting $\Delta_{ij}=0$, and the result for redistributing only $|\Delta_{ij}|$ mismatches derived before is reproduced for $k=0$.

Since the total number of mismatches between linked nodes can not exceed $m$, $l$ and $k$ are constrained by $l+2k+|\Delta_{ij}|\leqslant m$.
Summarizing we obtain for $\Delta_{ij}\geqslant0$
\begin{align}
  L_{ij} = \sum_{l, k=0} & {d\!-\!d_M \choose l} {i\!-\!1 \choose k}
  {d_M\!-\!i\!+\!1 \choose \Delta_{ij}+k} \nonumber \\ & \times
  \openone \left( l\!+\!2k\!+\!\Delta_{ij} \leqslant m
  \right) \,, \label{eq:linkmatrixii}
\end{align}
and for $\Delta_{ij}\leqslant0$
\begin{align} 
  L_{ij} = \sum_{l, k=0} & {d\!-\!d_M \choose l} {i\!-\!1 \choose
    |\Delta_{ij}|+k} {d_M\!-\!i\!+\!1 \choose k} \nonumber \\ & \times
  \openone \left( l\!+\!2k\!+\!|\Delta_{ij}| \leqslant m
  \right) \,. \label{eq:linkmatrixiii}
\end{align}
In some cases the evaluation of the sums in Eqs.~(\ref{eq:linkmatrixii}, \ref{eq:linkmatrixiii}) leads to simple rules, see 
\footnote{
For example, let $d=d_M\!+\!1$ and $m=2$. $\Delta_{ij} = \mbox{const} = 0, \pm 1, \pm 2$ for the different secondary diagonals, cf. Fig.~\ref{fig:lmstructure}.
For the middle secondary diagonal we have $\Delta_{ij} = 0$, i.e. $j=d_M\!-\!i\!+\!2$. Then, 
$
  L_{i,d_M-i+2}= 2+(i\!-\!1)(d_M\!+\!1\!-\!i)\,,
$
where $i=1, \dots, d_M\!+\!1\,$. The first secondary off-diagonals are specified by $\Delta_{ij} = 1$ or $-1$. Their entries are given by
$
  L_{i,d_M-i+1}=L_{d_M-i+2,i+1}=2(d_M\!+\!1\!-\!i)\,,
$
where   $i=1, \dots, d_M\,$. Similarly, the second secondary off-diagonals are specified by
$\Delta_{ij} = 2$ or $-2$. Their entries are
$
  L_{i,d_M-i}=L_{d_M-i+1,i+2}=(d_M\!+\!1\!-\!i)(d_M\!-\!i)/2 \,,
$
where $i=1, \dots, d_M\!-\!1\,$.
}.

Of course, the total number of links of a node is
\begin{equation}
\sum \nolimits_{j=1}^{d_M+1} L_{ij} = \kappa\,.
\end{equation}
The link matrix $\mathbb{L}=(L_{ij})$ has a symmetry given by
\begin{equation}
  L_{ij} = L_{d_M\!+\!2\!-\!i, d_M\!+\!2\!-\!j}\,. \label{eq:linkmatrix_symmetry}
\end{equation}
If the link matrix entries of the $i$th row are multiplied by the size
of group $S_i$, the resulting matrix with entries $\Lambda_{ij} =
|S_i| L_{ij}$ additionally obeys $\Lambda_{ij}=\Lambda_{ji}$.

It may be of interest to know the number of links from a node of $S_i$
to nodes of $S_j$ with a given number of mismatches $\mu$, which is
denoted by $L_{ij}^{\mu}$. It is obtained replacing in Eqs.~(\ref{eq:linkmatrixii}, \ref{eq:linkmatrixiii})
$\openone(l\!+\!2k\!+\!|\Delta_{ij}| \leqslant m)$ by $\delta_{
  l\!+\!2k\!+\!|\Delta_{ij}|,\mu}\,$, which gives, e.g. for $\Delta_{ij}\geqslant 0$
\begin{equation}
  L_{ij}^{\mu} = \sum_k {d - d_M \choose
    \mu\!-\!2k\!-\!\Delta_{ij}} {i\!-\!1 \choose k} 
  {d_M\!-i\!+\!1 \choose \Delta_{ij}\!+\!k}\,.
\end{equation}
Obviously, it holds $L_{ij} = \sum_{\mu=0}^{m} L_{ij}^{\mu}$.

\subsubsection{Architecture\label{sec:architecture}}

\begin{figure}[t]
\centering
\includegraphics[width=0.9\columnwidth]{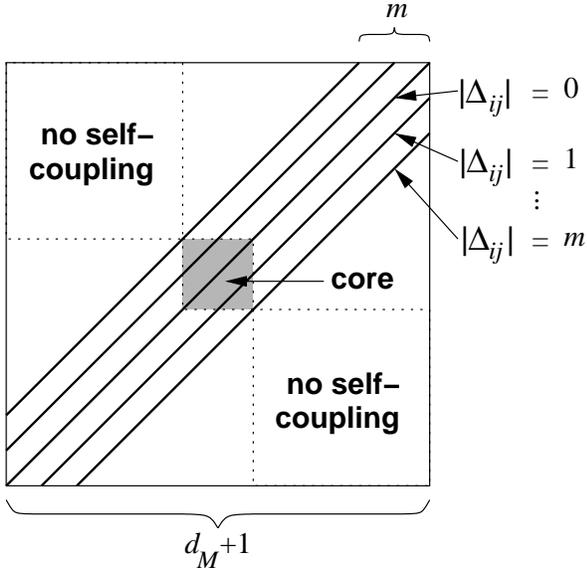}
\caption{The general structure of the $(d_M\!+\!1)\times(d_M\!+\!1)$ link matrix. Matrix entries along the $2m+1$ solid diagonal lines are non-zero, cf. Eqs.~(\ref{eq:linkmatrixii}, \ref{eq:linkmatrixiii}). All other entries are zero. Groups in the center (\textit{gray square}) couple to themselves and are naturally called core groups. Among the groups without self-coupling we can further distinguish groups that couple to the core from those which do not. See further discussion in the text.}
 \label{fig:lmstructure}
\end{figure}

For the link matrix and, thus, for the architecture of patterns we can state a number of general properties. Because $|\Delta_{ij}| = |d_M\!-\!i\!-\!j\!+\!2| \leqslant m$ is the necessary and sufficient condition for the existence of links between groups $i$ and $j$, all non-zero matrix entries are situated on a band along the secondary diagonal, see the illustration of the general structure of the link matrix in Fig.~\ref{fig:lmstructure}. The width of this band is $2m+1$, i.e. $m$ controls the range of influence of a group. Group $S_i$ interacts with the $2m+1$ groups $S_{d_M+2-i-m}\,, \dots\,, S_{d_M+2-i+m}$.

We always find groups with self-coupling, because there is always a
quadratic block of non-zero matrix entries in the center of
$\mathbb{L}$. Such groups are called core groups.  The size of the
core, i.e. the number of core groups, depends on $m$ and on the
existence of a central element in $\mathbb{L}$. Such an element exists
if $d_M$ is an even number.  Thus,
\begin{equation}
  \#\text{core groups} = \left\{ 
    \begin{array}{cc}
      m & \text{if either}\ m\ \text{or}\ d_M\ \text{odd}\\
      m\!+\!1 & \text{otherwise}
    \end{array}
  \right.\,.
\end{equation}

Groups without self-coupling exist if the number of groups $d_M\!+\!1$ is larger than the number of core groups. Such groups are related to the square submatrix of $\mathbb{L}$ with all entries equal to zero, cf. Fig.~\ref{fig:lmstructure}.

\begin{figure}[t]
\centering
\includegraphics[width=0.6\columnwidth]{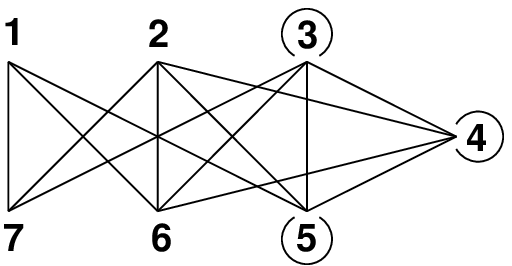}\\[10mm]
\includegraphics[width=0.9\columnwidth]{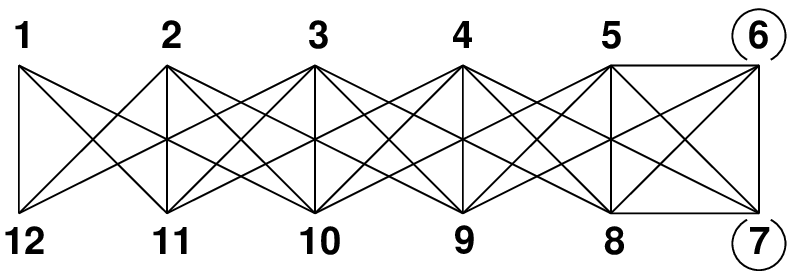}
\caption{Groups and their linking for even (above), and odd $d_M$
(below) on a two-mismatch base graph. The lines show possible links between nodes
of the groups. The circles indicate the self-coupling within the core
groups. It is suggestive to call arrangement of the groups caterpillar representation.}
\label{fig:Raupe} 
\end{figure}

The number of core groups for a given $m$ only depends on whether $d_M$ is even or odd but is independent on $d$. 
We can represent the content of Fig.~\ref{fig:lmstructure} in an alternative way exploiting that groups with index $i$ have the same properties as those with index $d_M+2-i$. 
In Fig.~\ref{fig:Raupe} these groups are on the same position in the two parallel strands. The core groups are on the right end of the diagram. We distinguish two cases, with even and odd number of core groups, respectively, which form the ``head'' of a caterpillar. If we increase $d_M$ by two, this does not change, but only the ``tail'' of the caterpillar gains an additional segment. This observation can be used to determine the scaling of the group sizes for large $d$, cf. the appendix of this paper.

Note, that all these general properties in this section are structural
information about the patterns. They only depend on the parameters
$d$, $m$ and on the choice of a pattern module $d_M$.

\subsubsection{Center of Mass\label{sec:centerofmass}}

In the simulations time series of $2^d$ nodes are generated, e.g. to calculate the mean occupation. In order to identify patterns in real time it has proven useful to reduce this information by introducing -- in analogy to classical mechanics -- a center of mass vector in dimension $d$. We consider occupied nodes, $n(v)=1$, as unit point masses in a $d$-dimensional space $[-1,1]^d$. The position vector $\bm{r}(v)$ of a node $v$ encoded by the bit chain $\bm{b}_d \bm{b}_{d-1} \cdots \bm{b}_1$ with $\bm{b}_i \in \{0, 1\}$ has components $r_i(v) = 2 \bm{b}_i -1$ in this space.
The center of mass is defined as
\begin{equation}
 \bm{R} = \frac{1}{n(G)} \sum_v n(v) \bm{r}(v)\,.
\end{equation}
The definition of $\bm{r}(v)$ ensures symmetry with respect to $\bm{r}=0$, which implies that for any symmetrically occupied pattern, e.g. the completely occupied base graph, we have $\bm{R}=0$.

\begin{figure}[t]
\centering
\includegraphics[width=\columnwidth]{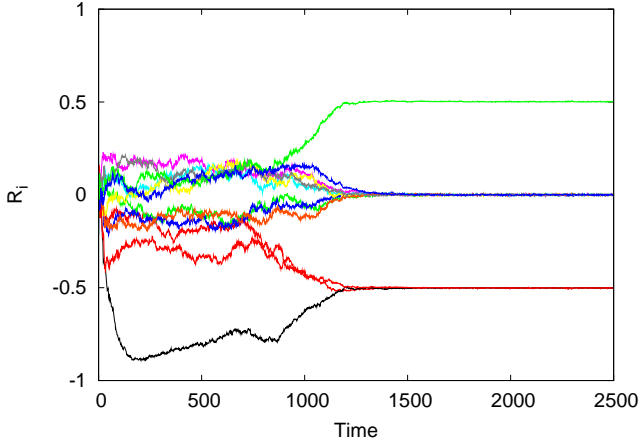}
\caption{(Color online) A typical time series of the center of mass vector components, here on $G_{12}^{(2)}$ with $[t_l, t_u]=[1,10]$ for $p=0.01$. The evolution starts from an empty base graph, which is gradually occupied thus breaking the symmetry. A stationary state is reached  after about 1300 time steps. Four components fluctuate around non-zero mean values, $R_1, R_7, R_{10} \approx -0.5$, and $R_{11}\approx 0.5$. Hence, as explained in the text, it is an architecture with $d_M=4$ determinant bits and all nodes in $S_1$ have $\bm{\cdot10\cdot\cdot0\cdot\cdot\cdot\cdot\cdot0}$. The time series of global quantities in Fig.~\ref{fig:timeseries} describes a different realization, which also evolves to a $d_M=4$ pattern.}
\label{fig:cootimeseries}
\end{figure}

Figure~\ref{fig:cootimeseries} is a typical example of a time series of the $d$ components of $\bm{R}$. Stationary states are characterized by small fluctuations of the components $R_i$ around some average values $\overline{R}_i$, think e.g. of a moving average. The value of $\overline{R}_i$ allows in the typical case to decide whether or not $i$ is a determinant bit position.

In general, for all non-determinant positions $i$ we have $\overline{R}_i\approx 0$. For any choice of determinant bits, the non-determinant bits run through all combinations of zeros and ones. Therefore, supposed that all nodes within a group are occupied with the same probability, the expected contribution of each group to the non-determinant components of $\bm{R}$ is zero.

For patterns which break the symmetry, the $R_i$ for determinant bit positions $i$ are non-zero, positive or negative. We explain in the following example, what can be inferred from this information.

\begin{figure}[t]
\begin{displaymath}
 \begin{array}{c@{\qquad}c}
  S_1 & \bm{1000}\\[0.15cm]
  S_2 & \bm{1001}\quad \bm{1010}\quad \bm{1100}\quad \bm{0000}\\[0.15cm]
  S_3 & \bm{1011}\quad \bm{1101}\quad \bm{0001}\quad \bm{1110}\quad \bm{0010}\quad \bm{0100}\\[0.15cm]
  S_4 & \bm{0110}\quad \bm{0101}\quad \bm{0011}\quad \bm{1111}\\[0.15cm]
  S_5 & \bm{0111}
 \end{array}
\end{displaymath}
\caption{For a pattern with $d_M=4$ we arrange the 4 determinant bits as they contribute to the 5 groups $S_1, \dots, S_5$ for the case that the determinant bits of $S_1$ are $\bm{1000}$, corresponding to Fig.~\ref{fig:cootimeseries}.}
\label{fig:coovector}
\end{figure} 

We consider a pattern with $d_M=4$. Fig.~\ref{fig:coovector} shows the determinant bits which contribute to the groups of this pattern. Non-determinant bits are not shown. In $S_2$ in each bit position the respective bits of $S_1$ predominate, in $S_4$ those of $S_5$. In $S_3$ the respective bits of $S_1$ and $S_5$ occur with the same frequency, cf. Fig.~\ref{fig:coovector}.

For a symmetry breaking pattern where $n(S_1\cup S_2) > n(S_4 \cup S_5)$, the sign of a determinant component $R_i$ is determined by the corresponding determinant bit $\bm{b}_i(v_1)$ of a node $v_1$ in $S_1$, $\mathrm{sign}\, R_i = r_i(v_1) = 2\bm{b}_i(v_1) - 1$. The other way round, measuring $\mathrm{sign}\, R_i$ we can infer $\bm{b}_i(v_1)$. If $n(S_1\cup S_2) < n(S_4 \cup S_5)$ we can return to the case above by relabeling the groups. The determinant bits $\bm{b}_i(v_1)$ of the pattern in Fig.~\ref{fig:cootimeseries} are $\bm{0}$ for $i=1, 7, 10$ and $\bm{1}$ for $i=11$.

The expectation value $\overline{R}_i$ for a given pattern is easily computed in terms of the expected occupation of the different groups $\overline{n}(S_j)$. For a given $j$ a fraction $(d_M-j+1)/d_M$ of the occupied nodes contributes $r_i(v_1)$, and a fraction ${(j-1)}/{d_M}$ contributes $-r_i(v_1)$. For a determinant bit at position $i$ we obtain
\begin{equation}
\overline{R}_i \approx \left[ \frac{1}{\overline{n}(G)} \sum_{j=1}^{d_M\!+\!1} \frac{d_M\!-\!2j\!+\!2}{d_M}\;\, \overline{n}(S_j) \right] r_i(v_1)\,, \label{eq:cooexpectation}
\end{equation} 
where we have supposed that the fluctuations of $n(G)$ are small. If $i$ is a non-determinant bit position, the arguments leading to Eq.~(\ref{eq:cooexpectation}) do not apply, but following a different line we obtain $\overline{R}_i \approx 0$ as explained above.

For the example shown in Fig.~\ref{fig:cootimeseries} we know from simulations that a static pattern occurs where all nodes of $S_2$ are occupied and the others empty, i.e. $n(G)=n(S_2)$. This leads to $\overline{R}_i \approx 0.5\, r_i(v_1)$.

For the typical case of a symmetry-breaking pattern, the $R_i$ allow to identify the determinant bits directly and in real time. This procedure is remarkably robust against defects in the pattern. 
For symmetric patterns, e.g. $n(S_1\cup S_2) = n(S_4 \cup S_5)$ in the above example, Eq.~(\ref{eq:cooexpectation}) gives $\overline{R}_i =0$, determinant and non-determinant bits can not be distinguished this way.

\section{Architecture of Specific Patterns\label{sec:specificpatterns}}

In the previous section we derived the structural properties. Here we apply these results to a zoo of specific patterns observed in simulations on $G_{12}^{(2)}$ with $[t_L,t_U]=[1,10]$. This restriction allows a certain completeness of the overview. For other parameters different patterns can be found, but inspection of several examples indicates that the building principles generally apply.

We start with the static patterns with $d_M=2, 4$ and $6$ and point out their common properties. We show how the concept of pattern modules needs to be extended to explain a more sophisticated static pattern. Finally, we describe the complex dynamic pattern which is build of modules of dimension $d_M=11$. 
For all examples below and a few more cases the link matrices are explicitly listed in the Supplementary Material \cite{STBsupplmat}.

\subsection{Simple Static Patterns\label{sec:staticpatterns}}


The simplest static pattern is the $d_M=2$ pattern with characteristic 2-clusters, discussed in detail in Sec.~\ref{sec:2cluster}. For the ideal pattern Eq.~(\ref{eq:cooexpectation}) gives $\overline{R}_i \approx 1\,r_i(v_1)$ for both determinant bit postions, all other $\overline{R}_i \approx 0$, which is in accordance with simulation.


In the static regime, for $0 < p \lessapprox 0.03$, we frequently observed a pattern with 8-clusters. It can be described with a pattern module of dimension $d_M\!=\!4$. 

\begin{figure}[t]
\centering
\includegraphics[width=0.95\columnwidth]{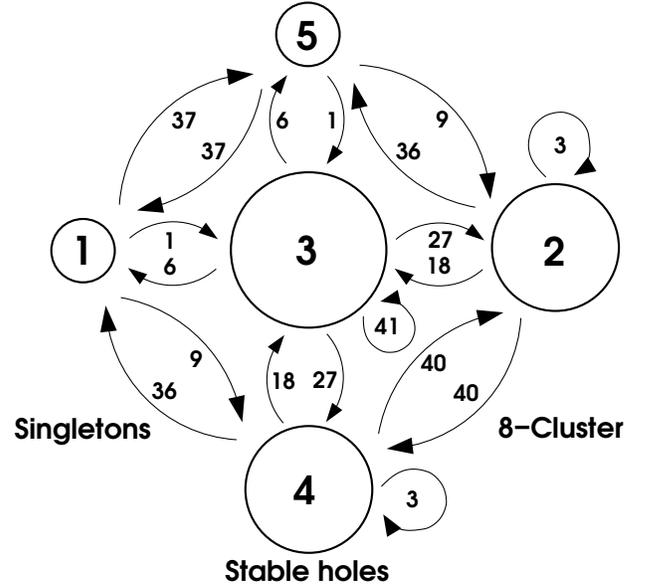}
\caption{Detailed view of the architecture of the 8-cluster pattern. As in Fig.~\ref{fig:scheme3g} the circle sizes correspond to the group sizes. The arrows and the numbers next to them indicate how many nodes of a group exert influence on the nodes of another group according to the link matrix.}
\label{fig:scheme5g}
\end{figure}

\begin{figure}[t]
\centering
\includegraphics[width=0.82\columnwidth]{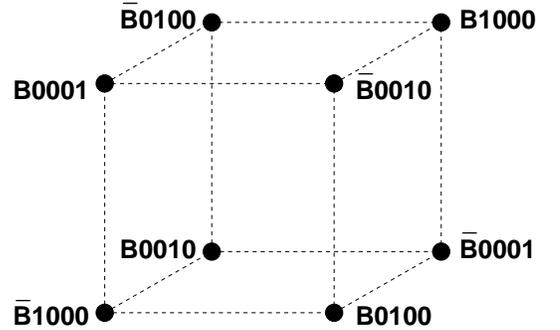}\\[2mm]
\caption{A cluster of eight occupied nodes as found in a snapshot of the occupied subgraph in the $d_M=4$ architecture. It has the topology of a cube in three dimensions. The cluster consists of nodes in $S_2$, which are all occupied up to few defects. The nodes are labelled with their bitstrings. The four digits represent the determinant bits. For simplicity and readability we give the determinant bits of the architecture realization in which $S_1$ has determinant bits $\bm{0000}$. $\bm{\mathsf{B}}$ represents the string of non-determinant bits, and $\overline{\bm{\mathsf{B}}}$ its inverse. All links have two mismatches.}
\label{fig:8cluster}
\end{figure}

It has five groups, which are illustrated in Fig.~\ref{fig:scheme5g} together with their links according to the link matrix, Eqs.~(\ref{eq:linkmatrixii},\ref{eq:linkmatrixiii}). Group $S_2$ is highly occupied and its nodes form the 8-clusters, cf. Fig.~\ref{fig:8cluster}. $S_3$, $S_4$ and $S_5$ are the groups of stable holes and $S_1$  is the group of singletons. Note the symmetry of the structure in the figure, by which we could swap the groups $S_2$ and $S_4$, and the groups $S_1$ and $S_5$. The same symmetry is reflected in the binomial coefficients in the formula for the group sizes, Eq.~(\ref{eq:groupsize}), and can be seen in the link matrix, cf. Eq.~(\ref{eq:linkmatrix_symmetry}).

The number of occupied nodes according to Eq. (\ref{eq:groupsize}) is $|S_2|={4 \choose 2-1} \times 2^8 = 1024$ and the number of singletons (unstable holes) is $|S_1|={4 \choose 1-1} \times 2^8=256$. The group of stable holes consists of 3 subgroups, the total number of stable holes is given by $|S_3|\!+\!|S_4|\!+\!|S_5| = (6+4+1) \times 2^8=2816$. This is in agreement with the observations given in Tab.~\ref{tab:globalq}.


For $0<p\lessapprox 0.03$ we find occasionally a $d_M=6$-pattern with characteristic 30-clusters. It is rare, but once established it remains stable for a long time. For the determinant bit positions in the ideal pattern Eq.~(\ref{eq:cooexpectation}) yields $\overline{R}_i \approx 0.33\,r_i(v_1)$ as in simulations.

\begin{figure}[t]
\centering
\includegraphics[width=0.6\columnwidth]{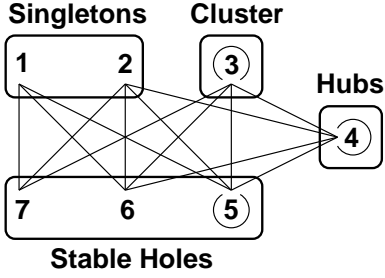}
\caption{The architecture of the 30-cluster pattern. As in Fig.~\ref{fig:Raupe} the lines show possible links between nodes of the groups.}
\label{fig:scheme7g}
\end{figure}

Its seven groups and their linking are schematically shown in Fig.~\ref{fig:scheme7g}. Each node in group $S_3$ has six links within the group. Thus, a completely occupied $S_3$ is stable. There are ${6 \choose 2}\!=\!15$ nodes in the pattern module which
belong to $S_3$. These are linked to the corresponding 15 nodes of the opposite module, thus  forming the 30-cluster. The stable hole groups $S_4$ to $S_7$ are suppressed, $S_1$ and $S_2$ are singletons.

These simple static patterns $d_M=2, 4, 6$ have in common a very regular structure and a mechanism of self-support and suppression. There is one fully occupied core group with a self-coupling within the threshold window $[t_L, t_U]=[1,10]$. Its nodes form clusters, sustain themselves and suppress all other groups except for the singletons, which are surrounded by stable holes only.

At $p=0$ the group of singletons will be emptied, the remaining occupied nodes which are connected to at least $t_L$ occupied nodes survive. An influx $p\gtrsim 0$ perturbs the pattern and tests its stability. Increasing $p$ further leads to more and more defects until, finally, the whole pattern will be destabilized.


All patterns observed in simulations can be well explained with the concept of pattern modules. On the contrary, not all patterns that can be constructed with pattern modules, have been actually observed. For many unobserved patterns we can explain why they are either forbidden or very rare.

By the pattern module $d_M=8$ we can construct a static 112-cluster pattern in the same manner as 2-, 8- and 30-cluster patterns. We occupy the core group $S_4$, which consists of clusters of size 112. Each node in such a cluster has 10 occupied neighbors, all within $S_4$. A small perturbation by occupying an additional neighbor destabilizes the pattern. This is the reason for its rareness. For even $d_M \geq 10$ the number of links within the core group exceeds $t_U$ so that a completely occupied core is impossible.

For odd $d_M$ the self-coupling of the core groups is so strong, $L_{ii}\geqslant12>t_U$, cf. \cite{STBsupplmat}, that static patterns with one completely occupied group are excluded. We could increase $t_U$ to allow some of these static patterns.

For our choice of $m=2$ in patterns with $d_M\geqslant7$ there are groups, at the end of the tail of the caterpillar, outside the range of influence of the core. An occupied core is not able to suppress these groups. The range of influence could be increased by increasing $m$.

\subsection{Static Pattern with Two Modules\label{sec:patternCombinedModules}}

There are static patterns that can not be explained with a single module, but by an extension of the concept to two pattern modules. We explain this with the following example.

\begin{figure}[t]
\centering
\includegraphics[width=\columnwidth]{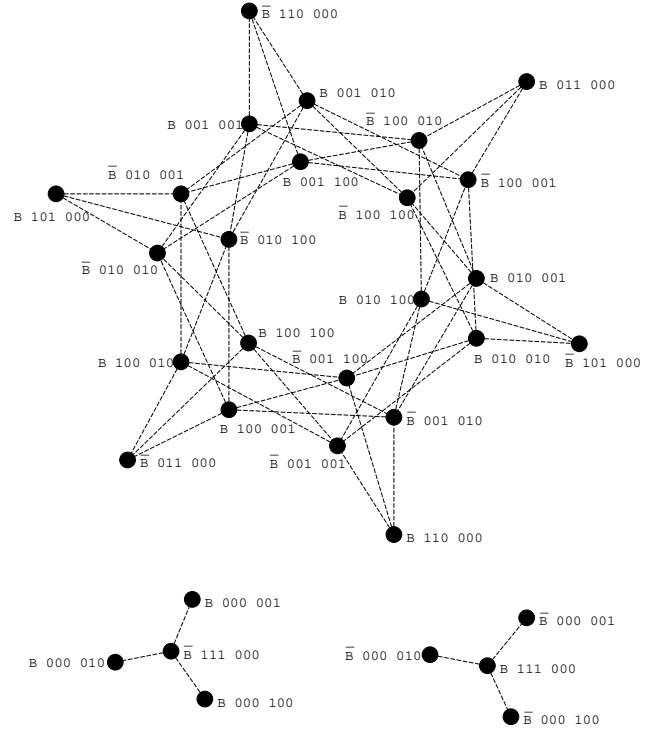}
\caption{A 24-cluster and two 4-clusters as observed in a snapshot of the occupied subgraph in the 2-module architecture with $d_M=d_M'=3$. The nodes are labeled with their bitstrings. The non-determinant bits are subsumed under $\bm{\mathsf B}$ or its inverse
$\overline{\bm{\mathsf B}}$, respectively. The two groups of three digits represent the determinant bits, corresponding to the first and second module. For readability the determinant bits are chosen such that group $S_1 \otimes S_1'$ is encoded by $\bm{000}$~$\bm{000}$. All links have two mismatches. Figure created using yEd \cite{yWorks}} \label{fig:2modules}
\end{figure}

In the regime of static patterns $0<p\lessapprox 0.035$, we often find a pattern that is characterized by large star-shaped clusters of size 24 and accompanying small clusters of
size four with one central node surrounded by three separate nodes attached to it, cf. Fig.~\ref{fig:2modules}.

It has 6 determinant bits, but the structure can not be explained by a single pattern module of dimension 6. However, an architecture constructed with two pattern modules of dimension $d_M$ and $d_M'$ with $d_M=d_M'=3$ is in full agreement with the observation. 

A node group is now denoted by $S_i \otimes S_j'$, the former corresponds to the first module, the latter to the second one. The group $S_1 \otimes S_1'$ has determinant bits $\bm{b}_1 \bm{b}_2 \bm{b}_3$~$\bm{b}'_1\bm{b}'_2\bm{b}'_3$, in the example given in Fig.~\ref{fig:2modules} we have chosen $\bm{000}$~$\bm{000}$. The index $k$ of $S_k$ has the same meaning as for a single module. The determinant bits of group $S_i \otimes S_j'$ deviate in $i-1$ bits from $\bm{b}_1 \bm{b}_2 \bm{b}_3$ and in $j-1$ bits from $\bm{b}'_1\bm{b}'_2\bm{b}'_3$. The deviations in the first set of bits are independent from those in the second set of bits. Therefore, not only 7 groups are generated as for a single module of $d_M=6$, but we find $(d_M+1)(d_M'+1) = 4\times4=16$ groups. The relative size of group $S_i \otimes S_j'$ is ${d_M \choose i-1}{d_M' \choose j-1}$. Also the link matrix can be calculated. The number of links from a node $v_{ij} \in S_i\otimes S_j'$ to nodes in $S_r\otimes S_s'$ is given by
\begin{align} 
 L_{ij,rs} & = \sum_{l,k,k'=0} \biggl[ {d\!-\!d_M\!-\!d_M' \choose l} \biggr. \\
	& \times {i - 1 \choose k \!+\! \max(0, -\Delta_{ij})} {d_M-i+1 \choose k \!+\! \max(0, \Delta_{ij})} \nonumber \\ 
	& \times {r - 1 \choose k' \!+\! \max(0, -\Delta'_{rs})} {d_M'-r+1 \choose k' \!+\! \max(0, \Delta'_{rs})} \nonumber \\ 
	& \times \biggl. \openone \Big( l\!+\!2k\!+\!2k'\!+ \!\left|\Delta_{ij}\right|\!+ \!\left|\Delta'_{rs}\right| \leqslant m \Big) \biggr] \nonumber \,,
\end{align}
where $\Delta'_{rs}=d_M'\!-\!r\!-\!s\!+\!2$ in analogy to $\Delta_{ij}$. This is a condensed notation for the four cases discriminated by $\Delta_{ij}, \Delta'_{rs} \gtrless 0$.

We obtain the ideal 24-clusters if we occupy the groups $S_2 \otimes S_2'$ for the nodes on the ring and $S_3 \otimes S_1'$ for the peripheral nodes.  Similarly the small 4-clusters are build of occupied groups $S_4 \otimes S_1'$ as the central node and $S_1 \otimes S_2'$ as the attached
nodes. In this static pattern more than one group is occupied. These groups mutually stimulate each other instead of supporting themselves.

For two modules the center of mass components $R_i$ differ not only between determinant and non-determinant bit positions, but also between determinant bits belonging to different modules. For the determinant bit at position $i$ Eq.~(\ref{eq:cooexpectation}) becomes
\begin{align}
  \overline{R}_i \approx & \left[ \frac{1}{\overline{n}(G)} \sum_{j=1}^{d_M\!+\!1} \sum_{j'=1}^{d_M'\!+\!1} \frac{d_M^{(\prime)}\!-\!2j^{(\prime)}\!+\!2}{d_M^{(\prime)}}\;\, \overline{n}\left( S_j\! \otimes\! S_{j'}' \right) \right] \nonumber \\
  & \times r_i(v_{1,1})\,,
\end{align} 
where $r_i(v_{1,1})$ is the position vector component of a node in $S_1 \otimes S_1'$ and the prime in parentheses at $d_M^{(\prime)}$ and $j^{(\prime)}$ only applies if $i$ is a position in the second module. For our pattern with $d_M=d_M'=3$ we observe $\overline{R}_i  \approx 0.29$ for the first module and $\overline{R}_i \approx 0.57$ for the second.

An extension to several modules of possibly different dimension appears natural.

\subsection{Dynamic Pattern}

In simulations starting from the empty base graph for $p\gtrapprox 0.03$ we only find a dynamical, stationary pattern with complex architecture. It was first observed in simulations in \cite{BB03}. There, six groups of nodes sharing statistical properties were identified and the architecture was described on a phenomenological base. There are groups of singletons, stable holes, two peripheral groups, and two core groups. A snapshot of the occupied nodes and their linking is given in Fig.~\ref{fig:gamma6g}.

\begin{figure}[t]
\centering
\includegraphics[width=0.9\columnwidth]{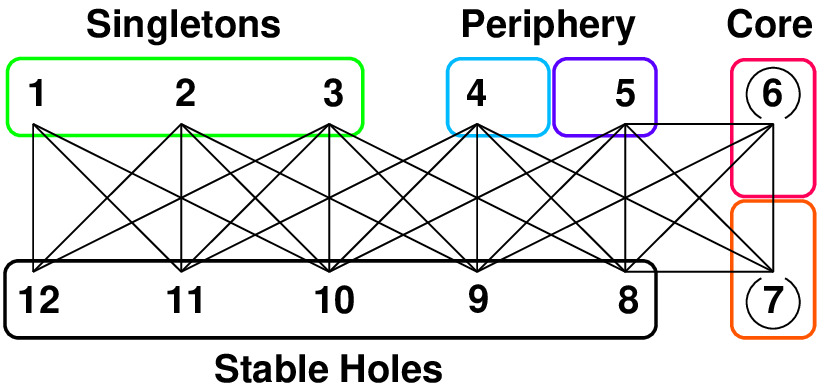}\\[5mm]
\includegraphics[width=0.9\columnwidth]{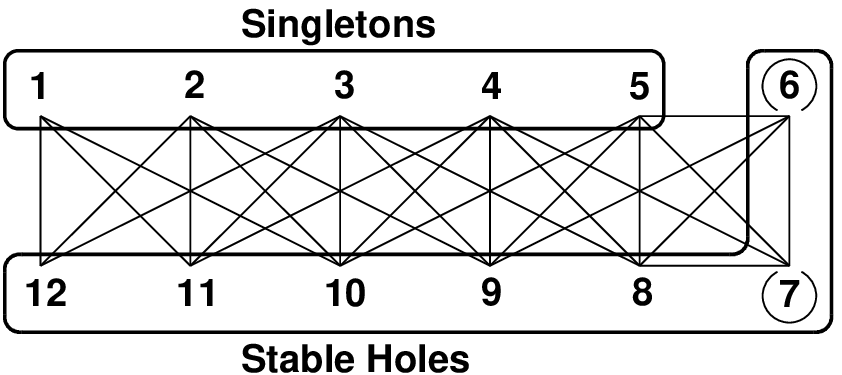}
\caption{(Color online) Visualization of the 12-group structure for $0.03\lessapprox p \lessapprox 0.045$ (top) and for $p\gtrapprox 0.045$ (bottom). As in Fig.~\ref{fig:Raupe} the lines show possible links between nodes of the groups. The coloring of the groups in the upper figure corresponds to the color of the respective nodes in Fig.~\ref{fig:gamma6g}, according to the qualitative classification.}
 \label{fig:scheme12gb}
\end{figure}

All structural properties, namely the number and size of the groups and their linking, can be explained within the concept of pattern modules using pattern modules of $d_M=11$. This leads to twelve groups, which for a certain range of $p$ can be merged to the six phenomenological groups above. Groups $S_1$ to $S_3$ are groups of singletons, and groups $S_8$ to $S_{12}$ are stable holes, see Fig.~\ref{fig:scheme12gb} (top) and Tab.~\ref{tab:module6g}. 

Table~\ref{tab:module6g} in the first two rows give the mapping from
the twelve groups $S_i$ to the six groups $\widetilde{S}_j$
found empirically in \cite{BB03}.  The derived group sizes $|S_i|$
are in excellent agreement with the measured group sizes. Also, the
sizes of the subgroups $S_8\,$, $S_9\,$, $S_{10}\,$, $S_{11}\,$, and
$S_{12}$ correctly sum up to 1124, which is exactly the
statistically measured number of stable holes, see Tab.~I in \cite{BB03}. 
Besides the structural information, Tab.~\ref{tab:module6g} also shows group averages of
local node characteristics, such as mean life time and mean occupation. For the group occupations $\langle \overline{n}(v) \rangle_{S_i}$ for $p=0.028$ given in Tab.~\ref{tab:module6g} we can determine $R_i$, for the determinant bits Eq.~(\ref{eq:cooexpectation}) and direct observation give $\overline{R}_i \approx 0.29$.

\begin{table*}[t]
\caption{The node characteristics of the 12-groups structure. Data from 500,000 iterations for $p\!=\!0.028$.\label{tab:module6g}\\ \vspace{2mm}}
\begin{ruledtabular}
\begin{tabular}{l*{12}{@{\quad}c}}
  group & $S_1$ & $S_2$ & $S_3$ & $S_4$ & $S_5$ & $S_6$ & $S_7$ & $S_8$ &
  $S_9$ & $S_{10}$ & $S_{11}$ & $S_{12}$ \\[1mm]
  \hline\\[-2mm]
  phenomenological group \cite{BB03} & $\widetilde{S}_4$ & $\widetilde{S}_4$ &
  $\widetilde{S}_4$ & $\widetilde{S}_5$ & $\widetilde{S}_6$ & 
  $\widetilde{S}_3$ & $\widetilde{S}_2$ & $\widetilde{S}_1$ &
  $\widetilde{S}_1$ & $\widetilde{S}_1$ & $\widetilde{S}_1$ &
  $\widetilde{S}_1$ \\[1mm]
  \hline \\[-2mm]
  qualitative classification & 
  \multicolumn{3}{|c|}{Singletons} &
  \multicolumn{2}{c|}{Periphery} &
  \multicolumn{2}{c|}{Core} &
  \multicolumn{5}{c|}{Stable holes} \\
  \\[-3mm] \hline\\[-2.5mm]
  group size $|S_i|$ & 2 & 22 & 110 & 330 & 660 & 924 & 924 & 660 & 330 & 110 & 22
  & 2 \\
  mean occupation $\langle \overline{n}(v) \rangle_{S_i}$ & 0.206 &	0.193 &	0.193 &	0.321 &	0.477 &	0.069
  &	0.030 &	0.000 &	0.000 &	0.000 &	0.000 &	0.002 \\
  mean life time $\langle \overline{\tau}(v)\rangle_{S_i}$&	9.11 &	8.53 &	8.56 &	16.86 &	32.55 &	2.66 &
  1.10 &	0.00 &	0.00 &	0.00 &	0.00 &	0.07 \\ 
  occupied neighbors $\langle \overline{n}(\partial v)\rangle_{S_i}$ &	0.01 &	0.01 &	0.01 &	0.84 &	1.88 &	8.94 &
  10.56 &	19.63 &	19.66 &	27.78 &	21.05 &	15.27 \\[1mm]
\end{tabular}
\end{ruledtabular}
\end{table*}

We calculated the link matrix for this pattern, cf.
Table~\ref{tab:links12g1}. In contrast to the static patterns that
emerge for low influx $p$ in this structure we also find perfect
matches and 1-mismatch links, but they are simply outnumbered by the
2-mismatch links.

\begin{table}[h]
\centering
\caption{Linkmatrix for $d_M\!=\!11$. Missing entries are zero. This is an instance for the general structure shown in Fig.~\ref{fig:lmstructure}.\label{tab:links12g1}\\
}
\begin{tabular}{|r|*{12}{c}|}
  \hline
  & $S_{1}$ & $S_{2}$ & $S_{3}$ & $S_{4}$ & $S_{5}$ & $S_{6}$ & $S_{7}$
  & $S_{8}$ & $S_{9}$ & $S_{10}$ & $S_{11}$ & $S_{12}$ \\ 
  \hline
  $v_{1}$ & &  &  &  &  &  &  &  &  &  55 & 22& 2 \\
  $v_{2}$ & &  &  &  &  &  &  &  &  45& 20 & 12& 2 \\ 
  $v_{3}$ & &  &  &  &  &  &  &  36& 18& 20 & 4 & 1 \\ 
  $v_{4}$ & &  &  &  &  &  &  28& 16& 26& 6& 3 &  \\
  $v_{5}$ & &  &  &  &  &  21& 14& 30& 8&  6&  &  \\  
  $v_{6}$ & &  &  &  &  15& 12& 32& 10& 10& &  &  \\ 
  $v_{7}$ & &  &  &  10& 10& 32& 12& 15& &  &  &  \\ 
  $v_{8}$ & &  &  6&  8&  30& 14& 21& &  &  &  &  \\ 
  $v_{9}$ & &  3&  6&  26& 16& 28& &  &  &  &  &  \\  
  $v_{10}$ & 1&  4&  20& 18& 36& & &  &  &  &  &  \\  
  $v_{11}$ & 2&  12& 20& 45& &  & &  &  &  &  &  \\  
  $v_{12}$ & 2&  22& 55& &  &  &  &  &  &  &  &  \\
  \hline
\end{tabular}
\end{table}

The structural properties and the mean occupation of the groups obtained in simulations allow to understand the qualitative behavior. The phenomenological classification in holes, singletons, etc. depends on $p$. 

For a range $0.03 \lessapprox p \lessapprox 0.045$ the following qualitative groups appear. There are \textit{stable holes} as in static patterns. \textit{Singletons} are surrounded by stable holes. \textit{Core groups} couple to all groups except for singletons. The number of self-couplings is larger than the upper threshold $t_U$. Therefore a complete occupation is not stable. However, there are many configurations of partial occupation. The \textit{periphery} couples to the core groups and to stable holes. It is stimulated by momentarily, during the influx occupied stable holes and by the stationary occupied core. It is the group with the highest occupation. The core groups have more links to occupied groups than the periphery and therefore its mean occupation is smaller than that of the periphery.

Although the pattern only evolves for $p\gtrapprox0.03$, the connected part of the network, the giant cluster, survives if the influx is stopped. Since $t_L=1$, all isolated nodes disappear. Starting with a small $p\lessapprox0.03$ from the giant cluster, leads to a similar scenario as described in Sec.~\ref{sec:randomevolution} where the giant cluster decays. This shows that the dynamic pattern requires a sufficiently high influx to emerge and to remain stationary. The influx permanently tests the stability of this pattern against random perturbations.

Increasing the influx above $p\gtrapprox 0.045$ also the core groups are suppressed by a
growing population of peripheral nodes. The occupation of the core
groups thus converges to the occupation of the stable holes while in
the same process the occupation of peripheral groups and singletons
converges. Thus, based on the same pattern module as above we can distinguish considering mean occupation and life time only two groups, stable holes and singletons, cf. Fig.~\ref{fig:scheme12gb} (bottom). Note, that taking the number of occupied neighbors into account we can still distinguish groups $S_6,\ \dots,\ S_9$.

For even higher $p\gtrapprox 0.07$ the dynamic pattern becomes transient. It often breaks down and rebuilds with a different orientation on the base graph. Nodes can no longer be permanently assigned to one of the twelve groups, but they change their group membership with every new formation of the pattern. Thus, a statistical characterization of nodes by temporal averages is impossible.

For still larger $p \gtrapprox t_U/\kappa = 0.127$ the dynamics is completely dominated by the random influx. The graph is so densely occupied after the influx that typically the upper threshold of the window rule is exceeded.

For $d_M \geqslant 7$ holes at the tail of the caterpillar are not sufficiently suppressed by the core, cf. the discussion at the end of Sec.~\ref{sec:staticpatterns} and 
Fig.~\ref{fig:Raupe}. Suppression by the periphery, $d_M\geqslant 7$, and by singletons, $d_M\geqslant 11$, is required. Otherwise these holes could become occupied, which would destabilize the complete pattern. Since singletons need sufficient influx, those patterns  occur only for higher $p$. Interestingly, among several candidates only the dynamic pattern with $d_M=11$ has been observed on $G_{12}^{(2)}$ for our setting of $[t_L,t_U]$ and $0<p<0.11$. It is also the only observed pattern with odd module dimension.

In this context we recall that the autonomous dynamics of the idiotypic network was a truly central feature of Jerne's original concept \cite{Jerne74}. The existence of core and periphery fits perfectly to the second generation idiotypic networks of Varela and Coutinho \cite{VC91}.

\section{Conclusions\label{sec:conclusions}}


We considered a minimalistic model \cite{BB03} of
the idiotypic network which evolves towards a complex functional architecture. The main mechanisms are the random influx of new idiotypes and the selection of not sufficiently stimulated idiotypes.  Numerical
simulations have shown that after a transient period a steady state with a specific  architecture is reached. Generally, groups of nodes sharing statistical properties
can be identified, which are linked in a characteristic way.

In the present paper we achieved a detailed analytical understanding of the building
principles of the emerging patterns. Modules of remarkable regularity serve as building blocks. We can calculate size and connectivity of the groups in agreement with numerical simulations.

The described building principles are formulated generalizing regularities found by a careful analysis of the simplest pattern. The architectures of all patterns observed so far can be described by these simple building principles. 

For a suitable parameter setting the network consists of a central and a peripheral part, as envisaged in the concept of second generation idiotypic networks \cite{VC91}. The central part is thought to play an essential role, e.g. in the control of autoreactive clones, the peripheral part to provide the response to external antigens and to keep a localized memory. An \emph{ad hoc} architecture similar to the one described here was used in \cite{SvHB94} to investigate the role of the idiotypic network in autoimmunity.

While the model is clearly immunologically inspired, it is worth pointing out that the results are interesting in themselves.
The process of self-ordering is genuinely evolutionary. The local processes of addition and deletion of nodes involve randomness (random influx) and selection (window rule),  they have a global impact. On a longer time scale we observe an evolution towards a complex architecture. Depending on the parameters this architecture can consist of many regularly arranged small modules or of few large modules covering the complete base graph.
Most interestingly, for a range of parameters, we find a dynamic stationary pattern comprising a central and a peripheral part. Also evolved real world networks like 
internet \cite{CHKSS07} and brain \cite{DCNSS06} exhibit central and peripheral parts.


The analytical understanding of the principles opens the possibility to consider networks of more realistic size and to investigate their scaling behavior, e.g. exploiting renormalization group techniques, cf. \cite{BB01}.

Future steps will include to check whether a similar understanding can
be reached for more realistic models.  For example, we think of
matching rules allowing bitstrings of different lengths, of links of
different weight for varying binding affinities, of several degrees of
population for each idiotype, of hypermutation during cell proliferation \cite{[{For a NK-type model of affinity maturation see }] DL03}, and of a delay of deletion of unstimulated clones. We also want to study the changes in the architecture during the life time of an individual.

First results of simulations on base graphs with varying link weights and several occupation levels indicate that the same building principles apply. Patterns with more than one pattern module appear to be more frequent \cite{Werner10}.

Ongoing studies investigate the evolution of the network in presence of self-antigen or an invading foreign antigen in terms of whether the network tolerates or rejects them \cite{Werner10,Willner11}.


In a forthcoming article \cite{SB} we report on a modular mean field theory to compute statistical group characteristics for arbitrary parameters and any given pattern. The mean occupation, the mean life time and the mean occupation of the neighborhood are in good agreement with the simulations.

\appendix*

\section{Scaling of the Relative Group Size with Increasing System Size\label{sec:scaling}}

Suppose the biological system can express about $10^{11}$ different genotypes for antibodies \cite{BM88}. This would correspond to a bitstring length $d=36$.
The system size in simulations is limited by the computer resources.
Still, today the biological system size is hard to reach. On the other hand, the thermodynamic limit is not of interest, because the biological system is large but finite.
Therefore, the scaling of qualitative properties with increasing system size is important.

In the following we consider patterns with odd $d_M=d\!-\!1$. For $d_M\! \geqslant\! 7$ we find from the ``caterpillar representation'', cf. Fig.~\ref{fig:Raupe}, an architecture of two core groups, two peripheral groups, and several groups of stable holes and of singletons. With increasing $d$ only the tail of the caterpillar grows, i.e. the number of stable hole and singleton groups increases. How does the relative size of the different qualitative groups scale with $d$?

The size of the groups is given by Eq.~(\ref{eq:groupsize}) which reduces for $d_M=d-1$ to 
\begin{equation}
 |S_i| = 2 {d\!-\!1 \choose i\!-\!1}\,,\ i=1,\dots,d\,.
\end{equation} 

We can order the groups with increasing $i$, then the two core groups are the two central groups for $i=d/2-1$ and $d/2$ which both have the same size. Thus the total size of the core groups is
\begin{equation}
 |C| = 2^2 {d\!-\!1 \choose d/2}\,.
\end{equation}

The peripheral groups are the two groups left of the core groups, $i=d/2-2$ and $d/2-3$, which have the total size 
\begin{equation}
 |P| = 2 \left[ {d-1 \choose d/2-2} + {d-1 \choose d/2-3}\right]\,.
\end{equation} 
Elementary algebra yields
\begin{equation}
 |P| = \frac{d(d-2)}{(d+2)(d+4)} |C|\,.
\end{equation}

Left to the peripheral groups are the groups of singletons, $S$, and right to the core groups are all groups of stable holes, $H$, cf. Table~\ref{tab:module6g} for the example with $d=12$. 

Because of ${n \choose m} = {n \choose n-m} $ we have
\begin{equation}
 |H| = |S| + |P|
\end{equation}
and obviously 
\begin{equation}
 |S| + |P| + |C| + |H| = |G_d| = 2^d\,.
\end{equation}
With this knowledge we infer
\begin{equation}
 |H| = 1/2 \left( |G_d| - |C| \right)
\end{equation}
and
\begin{equation}
 |S| = |H| - |P|\,.
\end{equation}

For large $d$ we obtain in Stirling approximation
\begin{equation}
 |C| \approx \sqrt{\frac{2}{\pi}} 2^{d+1} \frac{1}{\sqrt{d}} \left( 1 - \frac{3}{4 d^2} + \dots \right)\,.
\end{equation}
Thus, the relative size scales with $d$ as 
\begin{equation}
 |\widetilde{C}| = \frac{|C|}{|G_d|} \approx 2 \sqrt{\frac{2}{\pi}} \frac{1}{\sqrt{d}} \left( 1 - \frac{3}{4 d^2} \right) \sim d^{-1/2}\,.
\end{equation}

In the thermodynamic limit $|\widetilde{C}|$ and $|\widetilde{P}|$ tend to zero. For a realistic size of the network, e.g. $d=36$, we have for the qualitative groups
$ |\widetilde{C}| \approx 0.266$,
$ |\widetilde{P}| \approx 0.214$,
$ |\widetilde{H}| \approx 0.367$, and 
$ |\widetilde{S}| \approx 0.153$. $|\widetilde{C}|$ and $|\widetilde{P}|$ together account for a half of all nodes.

Of course, the parameters $p$ and $[t_L,t_U]$ have to be suitably adjusted such that a pattern of the supposed architecture actually emerges.

\begin{acknowledgments}
\hyphenation{IMPRS}
Thanks is due to Andreas K\"uhn, Heinz Sachsenweger, Benjamin Werner, and Sven Willner for valuable comments. H.S. thanks the IMPRS Mathematics in the Sciences and the Evangelisches Studienwerk Villigst e.V. for funding.
\end{acknowledgments}

\end{document}